\documentclass[aip,reprint,graphicx]{revtex4-1}

\draft 

\usepackage{amsbsy,amssymb,graphicx}
\usepackage{makeidx}\makeindex

\renewcommand{\bm}{\boldsymbol}

\newcommand{\dd}[2]{\frac{\partial{#2}}{\partial{#1}}}

\newcommand{\HALL}{\alpha}
\newcommand{\HEL}{\sigma}
\newcommand{\POL}{s}

\newcommand{\tripleHMHD}[4]{
  {#1(}\!{#1(} #2 {#1|}\!{#1|} #3 {#1|}\!{#1|} #4 {#1)}\!{#1)}
}

\newcommand{\Braket}[3]{{#1\langle}#2{#1|}#3{#1\rangle}}%

\newcommand{\U}{\underline}

\newcommand{\ddt}{\partial_{t}}

\newcommand{\LieBracket}[3]{{#1[}#2,#3{#1]}}
\newcommand{\curl}{\nabla\times}

\newcommand{\LieBraket}{\LieBracket}
\newcommand{\ADJ}[3]{{\rm{ad}}^{#1}_{#2}#3}

\newcommand{\Alfven}{Alfv\'en}
\newcommand{\M}{g}

\newcommand{\vs}{\vspace{1em}}
\newcommand{\sakujo}[1]{}

\begin{document}


\title{%
Particle-relabeling symmetry, generalized vorticity, and normal-mode expansion of ideal incompressible fluids and plasmas in three-dimensional space
} 



\author{Keisuke Araki}
 \email{araki@are.ous.ac.jp.\\ Tel: +81862569509. Fax: +81862553611.}
\affiliation{%
 Faculty of Engineering, Okayama University of Science,
 1-1 Ridai-cho, Kita-ku, Okayama 700-0005 JAPAN
}%

\date{\today}

\begin{abstract}
The Lagrangian mechanical consideration of the dynamics of ideal 
incompressible hydrodynamic, magnetohydrodynamic, 
and Hall magnetohydrodynamic media,
which are formulated as dynamical systems 
in appropriate Lie groups equipped with Riemannian metrics, 
leads to the notion of \textit{generalized vorticities}, 
as well as generalized coordinates, velocities, and momenta.
The action of each system is conserved against the integral path variation 
in the direction of the generalized vorticity, 
and
this invariance is associated with the particle relabeling symmetry.
The generalized vorticities are formulated by the operation of 
integro-differential operators on the generalized velocities.
The eigenfunctions of the operators provide sets of orthogonal functions,
and we obtain common mathematical expressions concerning these dynamical systems using the orthogonoal functions.
In particular, we find that 
the product of 
the Riemannian metric, $g_{im}$, 
and the structure constants of the Lie group, $C^{m}_{jk}$, 
is given by the product of 
the eigenvalue of the operator, $\Lambda(i)$, 
and a certain totally antisymmetric tensor, $T_{ijk}$: 
$g_{i\alpha}C^{\alpha}_{jk}=\Lambda(i)T_{ijk}.$
Its physical implications, including the weak interaction conjecture of MHD turbulence, are also discussed.
\end{abstract}

\pacs{52.30.-q,45.20.-d,52.35.Mw,47.10.-g}

\maketitle 


\section{Introduction}

Even though the operation of particle relabeling does not alter the fluid motions visible to us, the associated ``symmetry''\footnote%
{
Note that this symmetry is qualitatively different from the symmetry considered, for example, in gauge field theory, wherein group transformation is applicable in principle at any point in the relevant space and time (for an example, see Ref. \onlinecite{utiyama1956invariant}).
Conversely, particle relabeling symmetry implies that the fluid motion is determined irrespective of the initial configuration of the fluid labels and that the labels (i.e., the identity of each fluid parcel) at an assigned time are dependent on the history of the fluid or plasma motions.
Therefore, it should have been called ``initial label independence'' rather than ``relabeling symmetry''
}
is known to play a crucial role in considerations of the helicity conservation laws of hydrodynamic (HD)\cite{salmon1988hamiltonian} and magnetohydrodynamic (MHD)\cite{PadhyeMorrison1996} systems.

The degrees of freedom for the symmetry enter the description of the fluid mechanics when the Lagrangian specification is adopted.
%
The Lagrangian specification assigns a ``label'' to each fluid parcel to trace its motion and to distinguish it from other parcels.
%
The ``labeling'', therefore, enables the parcels to be distinguished.
However, the method of ``numbering'' the fluid parcels is arbitrary.
Despite this redundancy, the description of the fluid mechanics has a Lagrangian mechanical foundation.
\vs

The Lagrangian mechanical formulation is based on the introduction of a triplet (more generally an $n$-tuple on $n$-dimensional space) of functions ($\vec{X}$) that denotes the position of the fluid parcels.\footnote{%
In this paper, we place an arrow above a symbol to denote the multifunctional characteristic of the mathematical quantities. Boldface letters are used to denote vector fields.
}
As we will discuss in Section 2, because the composite of triplets $\vec{X}(\vec{Y})$ is also an element of the same function space, the space formally defines a continuous group.
This was recognized by Arnold when he reviewed his studies on dynamical systems in Lie groups equipped with a Riemannian metric and discussed the related hydrodynamic topics in a unified form.\cite{arnold1966sur}

As for plasma physics, extensions of the group theoretical formulation to ideal MHD systems were performed by Zeitlin and Kambe\cite{ZeitlinKambe1993} for the two-dimensional case and by Hattori\cite{hattori1994ideal} for the three-dimensional case.
The key concept that enables us to extend Arnold's formulation to multivariate variable systems is the \textit{semidirect product group}.\cite{HOLM19981}
The semidirect product of $G$ and $H$, $G \ltimes H$, is defined by the operation 
\begin{eqnarray}
(g_1,h_1) \circ (g_2,h_2) = (g_1 \circ g_2,h_1 \circ \rho(g_1)h_2),
\label{semidirect product}
\end{eqnarray}
where $\rho$ is a homomorphism of $G$ onto $H$.\cite{robinson2003introduction}
Many dynamical systems have been recognized as being in appropriate Lie groups.\cite{arnold1998topological,audin1999spinning}

For many cases of physical interest, the subsidiary group $H$ is often set to be a vector space, whose group operation is given by the addition of two elements.
However, this is a somewhat more limited situation than the definition given by Eq. (\ref{semidirect product}).
A general formulation, other than vector spaces, is found in Vizman's work.\cite{vizman2001geodesics}
This extension provides a way to formulate ideal incompressible Hall magnetohydrodynamics (HMHD) as a dynamical system in a semidirect product group of two diffeomorphism groups.\cite{araki2015differential}
(Another interesting finding is that the HMHD system is also formulated in a \textit{direct product group} of two diffeomorphism groups.\cite{araki2015helicity})

This approach has been applied to stability analyses of (possibly non-stationary) solutions of HD, MHD, and HMHD systems.
Group theoretic stability anlyses of homogeneous, isotropic MHD and HMHD turbulence conjecture that unstable interactions are limited to ``local'' interactions in wavenumber space, while non-local interactions are stable and have a ``wavy'' nature.\cite{araki2017differential2}
\vs


Investigations related to the helicity conservation laws have a very long history and contain a wide variety of topics,\cite{Moffatt3663} for example, their relationship to the topological nature of flow fields\cite{moffatt1969degree} and their influence on the dynamics of the plasma equilibrium state\cite{taylor1974relaxation} or turbulent phenomena.\cite{1992AnRFM..24..281M}

Of the numerous studies that have been conducted, helicity conservation, in particular, has been discussed from the viewpoint of relabeling symmetry by Salmon.\cite{salmon1988hamiltonian}
In this context, helicity is recognized to be one of the Casimir invariants.
This viewpoint was applied to an ideal MHD system by Padhye and Morrison.\cite{1996PlPhR..22..869P}
They discussed its relationship with Noether's theorem, which provides clear insights into the symmetry.
As a consequence of the application of the relabeling operation to the magnetic field, the magnetic helicity was derived, as well as the cross helicity.

As for the cases treated in the present study, particle-relabeling symmetry is simply formulated in terms of the \textit{Lin constraints}.\cite{marsden2013introduction}
These constraints describe the relationship between the reference and perturbed fields in terms of their relationship in an \textit{Eulerian} specification.
The Lin constraints require two types of perturbation fields. One is the label displacement field ($\bm\xi$), which expresses the perturbed position of each fluid parcel as $\vec{X}\to\vec{X}+\epsilon\bm{\xi}(\vec{X})$, where $\epsilon$ is a small parameter.
The other is the velocity perturbation field ($\bm v$), where the perturbation of the velocity field is given by $\bm{V}\to\bm{V}+\epsilon\bm{v}$.
As will be derived in Section 2, these two fields obey the equation
$$
  \bm{v} = \partial_{t}\bm{\xi} + [\bm{\xi},\bm{V}]
,
$$
where $[*,*]$ is the Lie bracket.
The second term on the right-hand-side is specific to dynamical systems in Lie groups and yields the quadratic terms of the evolution equation.
When $\bm{v}=\bm{0}$, or $\bm{\xi}$ satisfies the PDE
\begin{eqnarray}
  \partial_{t}\bm{\xi} + [\bm{\xi},\bm{V}] = \bm{0}
\label{relabelingCondition}
\end{eqnarray}
during a time interval (say $I=[0,1]$), the reference and perturbed paths give the same velocity history $\{\bm{V}(t);t\in I\}$.
Therefore, Eq. (\ref{relabelingCondition}) physically implies particle relabeling.
If the Lagrangian is solely defined by $\bm{V}$, the value of the Lagrangian and the action are invariant against the relabeling generated by the displacement history $\{\bm{\xi}(t);t\in I\}$.
Noether's theorem tells us that a corresponding conservation law exists.
Even though the choice of the initial condition $\bm{\xi}(0)$ is arbitrary, i.e., the degrees of freedom for this symmetry are very large, only the helicity conservation laws have been found for incompressible fluids and plasmas.
\vs

In the present study, the particle-relabeling symmetry is recognized by Eq. (\ref{relabelingCondition}) and we examine the HD, MHD, and HMHD systems.
As is presented in Section 2, the concrete formulas of the Lie brackets differ for these three systems.
The physical implication of Eq. (\ref{relabelingCondition}) departs from simple ``particle relabeling'' because the auxiliary variables (the current field for MHD) do not indicate the mere ``position of some material''.
%
However, this extensive viewpoint allows us to treat the dynamics of ideal incompressible fluids and plasmas in a unified manner.
\vs

This paper is organized as follows.
%
Section 2 is devoted to the reviews of the mathematical basis and the variational calculations.
%
Then, in Section 3, derivations of the generalized vorticities are presented in a somewhat heuristic way for the HD, MHD, and HMHD systems.
In the course of these calculations, we derive the integro-differential operators that generate action-preserving perturbations from the generalized velocities.
We call these the helicity-based particle-relabeling operators (or simply the ``relabeling operators'' hereafter).
Section 4 contains the core of this study: the eigenvalues and corresponding eigenfunctions of the relabeling operators are obtained as normal-modes; then, normal-mode expansions of the basic quantities, formulas, and evolution equations are presented.
%
In Section 5, the effects of a uniform magnetic field and the Coriolis force on the MHD and HMHD systems are discussed with respect to the associated linear waves.
The normal-mode expansion of the evolution equation is also presented.
%
Physical implications of our findings are discussed in Section 6.

\section{Lagrangian mechanical formalism in Lie groups}

In this section, we review the Lagrangian mechanics in Lie groups, which are equipped with an appropriate inner product, and its application to HD, MHD, and HMHD systems.

\subsection{Lagrangian specification of fluid motions and its extension to plasmas}

The Lagrangian specification of a flow field introduces a triplet of functions to indicate the positions of fluid particles at an assigned time:
\begin{eqnarray}
&&
  \vec{X}=
  \vec{X}(\vec{a},t)
\nonumber
\\&&=
  (X^1(a^1,a^2,a^3,t),X^2(a^1,a^2,a^3,t),X^3(a^1,a^2,a^3,t))
,
\nonumber
\end{eqnarray}
which indicates the position of a fluid particle at a time $t$ initially located at $(a^1,a^2,a^3)\in M$, where $M$ is a fluid container (mathematically a smooth manifold).
These triplets function as the \textit{generalized coordinates} of the Lagrangian mechanics of a continuum in a three-dimensional space.
Mathematically, the function space of these triplets, $G$, is equivalent to bijective maps from $M$ onto itself.
Because their composition
\begin{eqnarray}
  \vec{X}_{1} \circ \vec{X}_{2}
  =
  \vec{X}_{1}(\vec{X}_{2})
\nonumber
\end{eqnarray}
is also the elements of $G$, the function space formally constitutes a continuous \textit{group} with respect to the function composition operation, which is called a volume-preserving diffeomorphism group, $S$Diff($M$).
This is the key to understanding the basis of the mathematical description.

For the MHD system, the configuration space is given by the semidirect product of $S$Diff($M$) and the function space of the vector fields $\mathfrak{X}(M)$, which is an Abelian group with respect to the addition operation.
The resulting group operation is given by
\begin{eqnarray}
&&
  \big(\vec{X}_{1},-\HALL\tau_{1}\bm{J}_{1}\big) \circ
  \big(\vec{X}_{2},-\HALL\tau_{2}\bm{J}_{2}\big)
 \nonumber\\&&\hspace{2em}
  =
  \big(
    \vec{X}_{1}(\vec{X}_{2}),
    -\HALL(\tau_{1}\bm{J}_{1} + \tau_{2}{\rm{Ad}}_{\vec{X}_{1}}\bm{J}_{2})
  \big)
,
\label{MHD:group action def}
\end{eqnarray}
where 
$\tau$s and $\bm{J}$s are time parameters and vector fields, respectively,
Ad denotes the adjoint representation of $S$Diff($M$) on 
$\mathfrak{X}$($M$), and $\HALL$ is a constant.
Note that the Ad${}_{\vec{X}}$ operation physically implies the advection of a flozen-in line-element by the flow induced by ${\vec{X}}$.

For the HMHD system, the group operation on the semidirect product of two volume-preserving diffeomorphisms is defined by 
\begin{eqnarray}
  \big(\vec{X}_{1},\vec{Y}_{1}\big) \circ
  \big(\vec{X}_{2},\vec{Y}_{2}\big)
  =
  \big(
    \vec{X}_{1}(\vec{X}_{2}),
    \vec{Y}_{1}(\vec{X}_{1}(\vec{Y}_{2}(\vec{X}_{1}^{-1})))
  \big)
,\hspace{1em}
\label{HMHD:group action def}
\end{eqnarray}
where $\vec{X}$ and $\vec{Y}$ are such maps that induce $\bm{V}$ and $-\HALL\bm{J}$ which appear in the next subsection, respectively.

Note that, even though the generalized velocities are derived from the time derivatives of the generalized coordinates, the derivatives are mathematically improper as a vector field because the arguments of the coefficients and bases are different from each other.
Therefore, we introduce a corresponding vector field with an Eulerian specification, $\bm{V}=V^i\dd{x^i}{}$, which is defined as follows:
\begin{eqnarray}
  \left(V^i(t)\dd{x^i}{}\right)_{\vec{x}=\vec{X}(\vec{a},t)}
  &:=&
  \dd{t}{X^i}(\vec{a},t)\left.\dd{x^i}{}\right|_{\vec{x}=\vec{X}(\vec{a},t)}
.
\nonumber
\end{eqnarray}
Hereafter, all the variables are described using the Eulerian specification.

\subsection{Generalized velocity, inner product, and generalized momentum}
In the present study, the velocity field ${\bm{V}}$
for the HD system, and the velocity and current fields 
${\vec{\bm{V}}}:={}^t\!({\bm{V}},-\HALL{\bm{J}})$
for the MHD and HMHD systems, are chosen as the generalized velocities, where the parameter $\HALL$
is an arbitrary constant for the MHD system and the Hall-term strength parameter for the HMHD system.

The inner products of the generalized velocities, which are called the \textit{Riemannian metrics} in differential topological terminology, are given by
\begin{eqnarray}
  \Braket{\big}{\bm{V}_{1}}{\bm{V}_{2}}
  &:=&
  \int
    {\bm{V}}_{1} \cdot {\bm{V}}_{2} \,
  {\rm{d}}^3\vec{x}
,
\hspace{1em}
\label{HD:riemannianmetric}
\end{eqnarray}
for the HD system, and
\begin{eqnarray}
  \Braket{\big}{{\vec{\bm{V}}}_{1}}{{\vec{\bm{V}}}_{2}}
  :=
  \int
  \Big(
    {\bm{V}}_{1}\cdot{\bm{V}}_{2}
    + {\bm{B}}_{1} \cdot {\bm{B}}_{2}
  \Big)
  {\rm{d}}^3\vec{x}
,
\label{MHD:riemannianmetric}
\end{eqnarray}
for the MHD and HMHD systems, where $\bm{B}$ 
is the magnetic field induced by the current field $\bm{J}$
($\bm{J}=\curl\bm{B}$, $\nabla\cdot\bm{B}=0$).
Each inner product is chosen to give its total energy.

The Lagrangian is defined using the inner products such that
$
  L
  =\frac12\Braket{\big}{\dot{\gamma}(t)}{\dot{\gamma}(t)}_{\gamma(t)}
  =\frac12\Braket{\big}{\bm{V}(t)}{\bm{V}(t)}
$, 
where $\gamma(t)$ is an integral path.
The generalized momenta are defined by the functional derivatives of the Lagrangian:\footnote{Hereafter, the underlining of a boldface letter denotes an element of the generalized momentum space.}
$\displaystyle
  {\U{\vec{\bm{M}}}}:=(\partial{L}/\partial{\vec{\bm{V}}})
,
$
and therefore the generalized momenta are given by
${\vec{\U{\bm{M}}}}=\U{\bm{M}}_V$, where
\begin{eqnarray}
  \U{\bm{M}}_{V} := \dd{\bm{V}}{L} = \bm{V}
\end{eqnarray}
for the HD system, and 
${\vec{\U{\bm{M}}}}=(\U{\bm{M}}_V,\U{\bm{M}}_J)$, where
\begin{eqnarray}
  \U{\bm{M}}_{V} := \dd{\bm{V}}{L} = \bm{V}
,\ 
  \U{\bm{M}}_{J} := \dd{(-\HALL\bm{J})}{L} = -\HALL^{-1}\bm{A}
\end{eqnarray}
for the MHD and HMHD systems, respectively.
Hereafter, $\bm{A}$ is the vector potential of ${\bm{B}}$ 
($\bm{B}=\curl\bm{A}$, $\nabla\cdot\bm{A}=0$).

The relationships between the generalized momenta and the generalized velocities are expressed as ${\vec{\U{\bm{M}}}}=\widehat{M}{\vec{\bm{V}}}$,
where $\widehat{M}$ is the inertia operator given by
\begin{eqnarray}
  \widehat{M} = I
\end{eqnarray}
for the HD system, where $I$ is the identity operator, and
\begin{eqnarray}
  \widehat{M}
  =
  \left(\begin{array}{cc}
    I & O \\ 0 & (\HALL \curl)^{-2}
  \end{array}\right)
\end{eqnarray}
for the MHD and HMHD systems.

\subsection{The Lie bracket}

The Lie brackets are a fundamental mathematical structure that describe serial ``advections'' by the generalized velocities. That is, a sufficiently short segment of a path (for example, A$\to$B in Fig. \ref{derivation of Lin constraints}) is approximated by the exponential map of the tangent vector, $\bm{V}$: $\gamma(t+\tau)\approx e^{\tau \bm{V}(t)}\circ\gamma(t)$, and its product, i.e., the serial advection is given by the Baker-Campbell-Hausdorff formula:
\begin{eqnarray}
&&
  e^{\tau_{1} {\bm{V}}_{1}} \circ e^{\tau_{2} {\bm{V}}_{2}}
\nonumber\\&&
  \approx
  \exp(
    \tau_{1} {\bm{V}}_{1}+\tau_{2} {\bm{V}}_{2}
    +
    \frac12 \tau_{1} \tau_{2} 
    \LieBracket{}{{\bm{V}}_{1}}{{\bm{V}}_{2}}
    +
    o(\tau)
  )
.
\nonumber
\end{eqnarray}
Therefore, the Lie brackets reflect the characteristics of the advection of each system.
As for the incompressible media treated in the present study, they are given by
\begin{eqnarray}
  \LieBracket{\big}{\bm{V}_{1}}{\bm{V}_{2}}
  &:=&
  \curl \big( {\bm{V}_{1}} \times {\bm{V}_{2}} \big)
\label{HD:liebracket}
\end{eqnarray}
for the HD system,\cite{arnold1966sur} 
\begin{eqnarray}
  \LieBracket{\big}{\vec{\bm{V}}_{1}}{\vec{\bm{V}}_{2}}
  &:=&
  \Big(
    \curl \big( {\bm{V}_{1}} \times {\bm{V}_{2}} \big)
  ,
   \nonumber\\&&
    - \HALL \curl \big(
      {\bm{V}_{1}} \times {\bm{J}_{2}}
      +
      {\bm{J}_{1}} \times {\bm{V}_{2}}
    \big)
  \Big)
,
\label{MHD:liebracket}
\end{eqnarray}
for the MHD system,\cite{hattori1994ideal} and 
\begin{eqnarray}
  \LieBracket{\big}{\vec{\bm{V}}_{1}}{\vec{\bm{V}}_{2}}
  &:=&
  \Big(
    \curl \big( {\bm{V}_{1}} \times {\bm{V}_{2}} \big)
  ,
    -\HALL \curl \big(
      {\bm{V}_{1}} \times {\bm{J}_{2}}
     \nonumber\\&&
      +
      {\bm{J}_{1}} \times {\bm{V}_{2}}
      -\HALL
      {\bm{J}_{1}} \times {\bm{J}_{2}}
    \big)
  \Big)
,
\hspace{1em}
\label{HMHD:liebracket}
\end{eqnarray}
for the HMHD system.\cite{araki2015differential}
Note that, in this study, we use the relation
$
  \curl \big( \bm{a} \times \bm{b} \big)
  =
  (b^\alpha\partial_\alpha a^\beta-a^\alpha\partial_\alpha b^\beta)\partial_\beta
,
$
which holds when $\bm{a}$ and $\bm{b}$ are divergence-free.

\subsection{The Lin constraints}
\begin{figure}
\begin{center}
\includegraphics[width=0.5\textwidth]{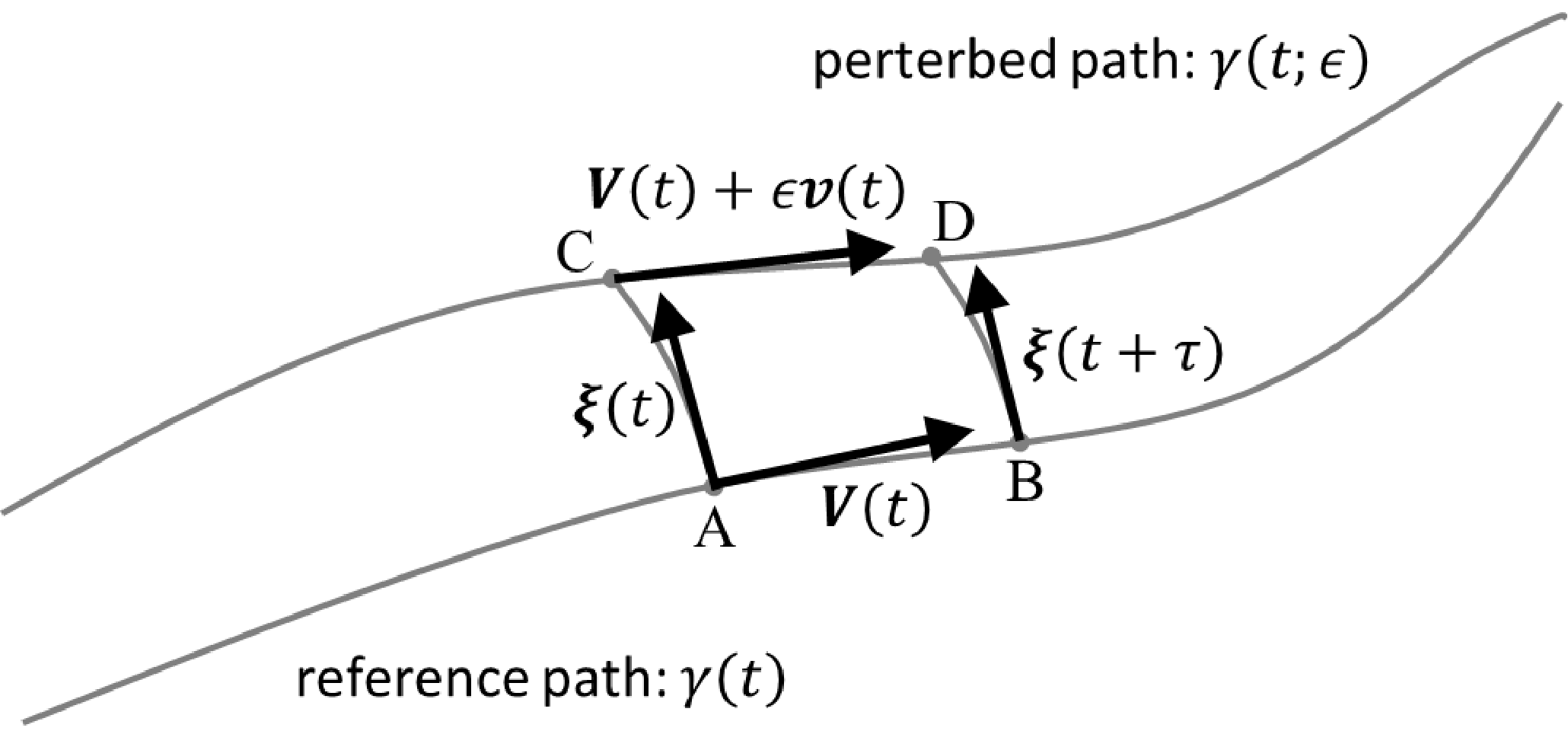}
\caption{\label{derivation of Lin constraints}Derivation of the Lin constraints,
i.e., the relationship between the velocity, the displacement, and the perturbation component of the velocity associated with the variation of a path.}
\end{center}
\end{figure}

Next, let $\gamma(t;\epsilon)$ be a perturbed path, where $\epsilon$ is a small perturbation parameter and $\epsilon=0$ corresponds to the reference path: $\gamma(t;0)=\gamma(t)$.
The path $\gamma(t;\epsilon)$ induces a label displacement field, $\bm{\xi}(t)$:
$
  \gamma(t;\epsilon)\approx e^{\epsilon\bm{\xi}(t)}\circ\gamma(t;0)
.
$
The variation of the path, $\gamma(t;\epsilon)$, induces a perturbation in the velocity, $\bm{v}$: i.e., approximating the small segment C$\to$D in Fig.
\ref{derivation of Lin constraints} via an exponential map, the point D is evaluated as
\begin{eqnarray}
  \gamma(t+\tau;\epsilon) \approx
    \exp[\tau(\bm{V}(t)+\epsilon\bm{v}(t))] \circ \gamma(t;\epsilon)
.
\label{CtoD}
\end{eqnarray}
Conversely, bypassing the reference path via C$\to$A$\to$B$\to$D in Fig. \ref{derivation of Lin constraints}, the point D is also evaluated as
\begin{eqnarray}
  \gamma(t+\tau;\epsilon) \approx
    e^{\epsilon\bm{\xi}(t+\tau)} \circ e^{\tau \bm{V}(t)} \circ e^{-\epsilon\bm{\xi}(t)}
    \circ \gamma(t;\epsilon)
.
\label{CtoAtoBtoD}
\end{eqnarray}
Applying the Baker-Campbell-Hausdorff formula to Eq. (\ref{CtoAtoBtoD}) and comparing the $O(\epsilon\tau)$ terms of Eqs. (\ref{CtoD}) and (\ref{CtoAtoBtoD}), we obtain the following relationship between $\bm{V}$, $\bm{\xi}$, and $\bm{v}$:
\begin{eqnarray}
  \bm{v}
  =
  \ddt{\bm{\xi}} + \LieBracket{\big}{\bm{\xi}}{\bm{V}}
,
\end{eqnarray}
which are known as the \textit{Lin constraints}.\cite{marsden2013introduction}
Corresponding to the Lie brackets, the Lin constraints are given by
\begin{eqnarray}
  \bm{v} & = &
  \ddt{\bm{\xi}} + \curl ( {\bm{\xi}} \times {\bm{V}} )
\label{HD:linconstraints1u}
\end{eqnarray}
for the HD system,
\begin{eqnarray}
  \bm{v} & = &
  \ddt{\bm{\xi}} + \curl ( {\bm{\xi}} \times {\bm{V}} )
,
\label{MHD:linconstraints1u}
\\
  \bm{j} & = &
  \ddt{\bm{\eta}}
  +
  \curl \big(
    {\bm{\xi}} \times {\bm{J}}
    +
    {\bm{\eta}} \times {\bm{V}}
  \big)
\label{MHD:linconstraints1j}
\end{eqnarray}
for the MHD system, and
\begin{eqnarray}
  \bm{v} & = &
  \ddt{\bm{\xi}} + \curl ( {\bm{\xi}} \times {\bm{V}} )
,
\label{HMHD:linconstraints1u}
\\
  \bm{j} & = &
  \ddt{\bm{\eta}}
  +
  \curl \big(
    {\bm{\xi}} \times {\bm{J}}
    +
    {\bm{\eta}} \times {\bm{V}}
    -
    \HALL {\bm{\eta}} \times {\bm{J}}
  \big)
\label{HMHD:linconstraints1j}
\end{eqnarray}
for the HMHD system.

\subsection{The Euler-Lagrange equation}
Therefore, the first variation of the action along a perturbed path,
$
  S_{\epsilon}=\int_{0}^{1}L
  (\gamma(t;{\epsilon}),\dot{\gamma}(t;{\epsilon}))\,
  {\rm{d}}t
,
$ 
becomes
\begin{eqnarray}
  \left.\dd{\epsilon}{S_{\epsilon}}\right|_{\epsilon=0}
  & = & 
  \int_{0}^{1}{\rm{d}}t
  \Braket{\big}{\vec{\bm{V}}}{\vec{\bm{v}}}
  = 
  \int_{0}^{1}{\rm{d}}t
  \Braket{\big}{
    \vec{\bm{V}}
  }{
    \ddt\vec{\bm{\xi}}+\LieBraket{\big}{\vec{\bm{\xi}}}{\vec{\bm{V}}}
  }
\nonumber\\
  & = & 
  \Braket{\big}{\vec{\bm{V}}}{\vec{\bm{\xi}}}\Big|^{t=1}_{t=0}
  -
  \int_{0}^{1}{\rm{d}}t
  \Braket{\big}{
    \ddt\vec{\bm{V}} - \ADJ{\dag}{\vec{\bm{V}}}{\vec{\bm{V}}}
  }{
    \vec{\bm{\xi}}
  }
,
\hspace{1em}
\label{the first variation}
\end{eqnarray}
where $\ADJ{\dag}{}{}$ is the dual operator of the Lie bracket with respect to the inner product:\footnote{
Note that the adjoint operator $\ADJ{\dag}{}{}$ is denoted by $B(*,*)$ in Arnold's work.\cite{arnold1966sur,arnold1998topological}
}%
\begin{eqnarray}
  \Braket{\big}{\ADJ{\dag}{\vec{\bm{c}}}{\vec{\bm{a}}}}{\vec{\bm{b}}}
  :=
  \Braket{\big}{\vec{\bm{a}}}{\LieBraket{\big}{\vec{\bm{b}}}{\vec{\bm{c}}}}
.
\label{def:ad dagger}
\end{eqnarray}

Applying Hamilton's principle to Eq. (\ref{the first variation}), 
more exactly, the reduced variational principle 
(Ref. \onlinecite{marsden2013introduction}, Sect. 13),
$(\partial S_{\epsilon}/\partial\epsilon)_{\epsilon=0}=0$, and $\vec{\bm{\xi}}=0$ at $t=0$, 1, 
we obtain the \textit{Euler-Poincare equation},
\begin{eqnarray}
  \ddt\vec{\bm{V}}-\ADJ{\dag}{\vec{\bm{V}}}{\vec{\bm{V}}}=0
,
\label{the euler-poincare eq}
\end{eqnarray}
as the Euler-Lagrange equation.

Corresponding to the Lin constraints, Eqs.
(\ref{HD:linconstraints1u})-(\ref{HMHD:linconstraints1j}), 
the Euler-Lagrange equations become 
\begin{eqnarray}
  \ddt{\bm{V}}
  + {\bm{\Omega}} \times {\bm{V}}
  & = & - \nabla P
\label{HD:eulerlagrangev}
\end{eqnarray}
for the HD system, 
\begin{eqnarray}
  \ddt{\bm{V}}
  + {\bm{\Omega}} \times {\bm{V}}
  + {\bm{B}} \times {\bm{J}}
  & = & - \nabla P
,
\label{MHD:eulerlagrangev}
\\
  \ddt{\bm{A}}
  + {\bm{B}} \times {\bm{V}}
  & = & - \nabla \phi
\label{MHD:eulerlagrangej}
\end{eqnarray}
for the MHD system, and
\begin{eqnarray}
  \ddt{\bm{V}}
  + {\bm{\Omega}} \times {\bm{V}}
  + {\bm{B}} \times {\bm{J}}
  & = & - \nabla P
,
\label{HMHD:eulerlagrangev}
\\
  \ddt{\bm{A}}
  + {\bm{B}} \times {\bm{V}}
  - \HALL {\bm{B}} \times {\bm{J}}
  & = & - \nabla \phi
\label{HMHD:eulerlagrangej}
\end{eqnarray}
for the HMHD system, where 
$\bm{\Omega}$, $P$, and $\phi$ are the vorticity ($\bm{\Omega}=\curl\bm{V}$), 
the generalized pressure, and the scalar potential, respectively.

\section{Generalization of the vorticity equation}

Equation (\ref{the first variation}) tells us that, once we find such a variable $\vec{\bm{\xi}}$ that satisfies 
\begin{eqnarray}
  \ddt\vec{\bm{\xi}} + \LieBracket{\big}{\vec{\bm{\xi}}}{\vec{\bm{V}}} = \vec{\bm{0}}
\label{advection}
\end{eqnarray}
along the solution path to the Euler-Lagrange equation, Eq.
(\ref{the euler-poincare eq}), 
we obtain the conservation law 
\begin{eqnarray}
  \Braket{\big}{\vec{\bm{V}}}{\vec{\bm{\xi}}}_{t=0}=\Braket{\big}{\vec{\bm{V}}}{\vec{\bm{\xi}}}_{t=1}
\label{noether}
\end{eqnarray}
as Noether's first theorem
(because the values of the Lagrangian, and therefore the action, are invariant against the variation of the integral path).

In the context of fluid and plasma physics, this invariance is called particle-relabeling symmetry because it changes the assigned value of each fluid particle in the Lagrangian specification without changing the velocity in the Eulerian specification.
%
%

In this section, we will see that Eq. (\ref{advection}) 
can be derived from the Euler-Lagrange equations for the HD, MHD, and HMHD systems.
The calculation processes themselves are a reconfirmation of the derivation of the helicity invariants.
However, the derivation procedures naturally define certain differential operators (denoted by $\widehat{D}$), which will be shown to provide clues to simplifying the normal-mode-expansion analysis of the dynamics.

\subsection{HD}
Taking the curl of Eq. (\ref{HD:eulerlagrangev}),
we obtain the vorticity equation:
\begin{eqnarray}
  \ddt{\bm{\Omega}}
  +
  \curl (
    {\bm{\Omega}} \times {\bm{V}}
  )
  & = & \bm{0}
.
\label{HD:eqofmotOv}
\end{eqnarray}
In terms of the Lie bracket, this equation is rewritten as
\begin{eqnarray}
  \ddt{\bm{\Omega}}
  +
  \LieBracket{\big}{\bm{\Omega}}{\bm{V}}
  =
  {\bm{0}}
.
\label{vorticity equation}
\end{eqnarray}
This equation obviously satisfies the particle-relabeling condition
(${\bm{v}}={\bm{0}}$ for Eq. (\ref{HD:linconstraints1u})).
The associated constant of motion, Eq. (\ref{noether}) is the helicity:\cite{moffatt1969degree}
\begin{eqnarray}
  H_{\rm{HD}} & = &
  \int
    {\bm{V}} \cdot {\bm{\Omega}}
  \,{\rm{d}}^3\vec{x}
.
\hspace{1em}
\end{eqnarray}
To discuss the general theory, we define 
\begin{eqnarray}
\widehat{D}:=\curl
\end{eqnarray}
for the HD system.

\subsection{MHD}
Taking the curl of Eqs. 
(\ref{MHD:eulerlagrangej}) and (\ref{MHD:eulerlagrangev}):
\begin{eqnarray}
  \ddt{\bm{B}}
  +
  \curl (
    {\bm{B}} \times {\bm{V}}
  )
  & = & \bm{0}
,
\label{MHD:eqofmotOj}
\\
  \ddt{\bm{\Omega}}
  +
  \curl (
    {\bm{\Omega}} \times {\bm{V}}
    +
    {\bm{B}} \times {\bm{J}}
  )
  & = & \bm{0}
,
\label{MHD:eqofmotOv}
\end{eqnarray}
rewriting Eq. (\ref{MHD:eqofmotOj}) as the following pair of equations:
\begin{eqnarray}
  \ddt{\bm{0}}
  +
  \curl (
    {\bm{0}} \times {\bm{V}}
  )
  & = & \bm{0}
,
\nonumber
\\
  \ddt{\bm{B}}
  +
  \curl (
    {\bm{0}} \times {\bm{J}}
    +
    {\bm{B}} \times {\bm{V}}
  )
  & = & \bm{0}
,
\label{MHD:generalized vorticity eq 2}
\end{eqnarray}
and
comparing Eqs. (\ref{MHD:eqofmotOj}), (\ref{MHD:eqofmotOv}), 
and (\ref{MHD:generalized vorticity eq 2})
to Eqs. (\ref{MHD:linconstraints1u}) and (\ref{MHD:linconstraints1j}),
we find that the variable,
\begin{eqnarray}
  {\vec{\bm{\Omega}}}
  =
  \left(\begin{array}{c}
    {\bm{\xi}}_{\Omega} \\ - \HALL {\bm{\eta}}_{\Omega}
  \end{array}\right)
  =
  C_{C}
  \left(\begin{array}{c}
    {\bm{B}}
  \\
    - \HALL {\bm{\Omega}}
  \end{array}\right)
  -
  {C_{M}}
  \left(\begin{array}{c}
    {\bm{0}}
  \\
    {\bm{B}}
  \end{array}\right)
,
\hspace{1em}
\label{MHD:generalizedVorticity}
\end{eqnarray}
(where $C_{C}$, and $C_{M}$ are arbitrary constants) 
satisfies the equation
\begin{eqnarray}
  \ddt{\vec{\bm{\Omega}}}
  +
  \LieBracket{\big}{\vec{\bm{\Omega}}}{\vec{\bm{V}}}
  =
  {\vec{\bm{0}}}
,
\label{generalized vorticity equation}
\end{eqnarray}
i.e., the particle-relabeling condition
($\bm{v}=\bm{j}=\bm{0}$ for Eqs. 
(\ref{MHD:linconstraints1u}) and (\ref{MHD:linconstraints1j})).
Because Eq. (\ref{generalized vorticity equation})
formally satisfies the same equation as Eq. (\ref{vorticity equation}),
we call the variable ${\vec{\bm{\Omega}}}$ 
the \textit{generalized vorticity}.
The derivation process is summarized as the operation of
the differential operator $\widehat{D}$ defined as
\begin{eqnarray}
  \widehat{D}
  & := &
  \left(\begin{array}{cc}
    O & - C_{C} \HALL \curl
  \\
    - C_{C} \HALL \curl & C_{M} \HALL \curl
  \end{array}\right)
,
\label{MHD:opD}
\end{eqnarray}
on the equation for the generalized momenta: 
${\vec{\bm{\Omega}}}=\widehat{D}{\vec{\U{\bm{M}}}}$.

The associated constants of motion, Eq. (\ref{noether}), 
are the linear combination of the cross and magnetic helicities:\cite{moffatt1969degree,woltjer1958theorem}
\begin{eqnarray}
  H_{\rm{MHD}} & = &
  2 C_{C} \int
    {\bm{V}} \cdot {\bm{B}}
  \,{\rm{d}}^3\vec{x}
  +
  \frac{C_{M}}{\HALL}
  \int
  {\bm{A}} \cdot {\bm{B}} \,{\rm{d}}^3\vec{x}
.
\label{MHD:derived helicity}
\hspace{1em}
\end{eqnarray}

\subsection{HMHD}
The curls of Eqs. (\ref{HMHD:eulerlagrangev}) and 
(\ref{HMHD:eulerlagrangej}) are
\begin{eqnarray}
  \ddt{\bm{\Omega}}
  +
  \curl (
    {\bm{\Omega}} \times {\bm{V}}
    +
    {\bm{B}} \times {\bm{J}}
  )
  & = & \bm{0}
,
\label{HMHD:eqofmotOv}
\\
  \ddt{\bm{B}}
  +
  \curl (
    {\bm{B}} \times {\bm{V}}
    -
    \HALL {\bm{B}} \times {\bm{J}}
  )
  & = & \bm{0}
.
\label{HMHD:eqofmotOj}
\end{eqnarray}
Calculating the combination
$\HALL\times$Eq. (\ref{HMHD:eqofmotOv}) + Eq. (\ref{HMHD:eqofmotOj}) 
and Eq. (\ref{HMHD:eqofmotOv}),
\begin{eqnarray}
  \ddt ( \HALL {\bm{\Omega}} + {\bm{B}} )
  + \curl [ ( \HALL {\bm{\Omega}} + {\bm{B}} ) \times {\bm{V}} ]
  &=& \bm{0}
,
\nonumber
\\
  \ddt{\bm{\Omega}}
  +
  \curl (
    ( \HALL {\bm{\Omega}} + {\bm{B}} ) \times {\bm{J}}
    +
    {\bm{\Omega}} \times {\bm{V}}
    -
    \HALL {\bm{\Omega}} \times {\bm{J}}
  )
  &=& \bm{0}
,
\nonumber
\\
\label{HMHD:generalized vorticity eq 1}
\end{eqnarray}
rewriting Eq. (\ref{HMHD:eqofmotOj}) 
as the following pair of equations:
\begin{eqnarray}
  \ddt{\bm{0}}
  +
  \curl (
    {\bm{0}} \times {\bm{V}}
  )
  & = & \bm{0}
,
\nonumber
\\
  \ddt{\bm{B}}
  +
  \curl (
    {\bm{0}} \times {\bm{J}}
    +
    {\bm{B}} \times {\bm{V}}
    -
    \HALL {\bm{B}} \times {\bm{J}}
  )
  & = & \bm{0}
,
\label{HMHD:generalized vorticity eq 2}
\end{eqnarray}
and comparing Eqs. (\ref{HMHD:generalized vorticity eq 1}) and 
(\ref{HMHD:generalized vorticity eq 2})
to Eqs. 
(\ref{HMHD:linconstraints1u}) and (\ref{HMHD:linconstraints1j}),
we find that the variable,
\begin{eqnarray}
  {\vec{\bm{\Omega}}}
  =
  \left(\begin{array}{c}
    {\bm{\xi}}_{\Omega} \\ - \HALL {\bm{\eta}}_{\Omega}
  \end{array}\right)
  =
  C_{C}
  \left(\begin{array}{c}
    \HALL {\bm{\Omega}} + {\bm{B}}
  \\
    - \HALL {\bm{\Omega}}
  \end{array}\right)
  -
  {C_{M}}
  \left(\begin{array}{c}
    {\bm{0}}
  \\
    {\bm{B}}
  \end{array}\right)
,
\hspace{1em}
\label{HMHD:generalized vorticity}
\end{eqnarray}
satisfies the particle-relabeling condition
($\bm{v}=\bm{j}=\bm{0}$ for Eqs. 
(\ref{HMHD:linconstraints1u}) and (\ref{HMHD:linconstraints1j})), 
where $C_{C}$, and $C_{M}$ are arbitrary constants.
This implies that, analogous to the MHD system, 
the generalized vorticity ${\vec{\bm{\Omega}}}$ 
formally satisfies the same equation as Eq. (\ref{generalized vorticity equation})
and the associated differential operator $\widehat{D}$
can be defined as follows:
\begin{eqnarray}
  \widehat{D}
  & := &
  \left(\begin{array}{cc}
    C_{C} \HALL \curl & - C_{C} \HALL \curl
  \\
    - C_{C} \HALL \curl & C_{M} \HALL \curl
  \end{array}\right)
.
\label{HMHD:opD}
\end{eqnarray}
Therefore, we call $\widehat{D}$ the generalized curl operator.

The associated constant of motion is given by
\begin{eqnarray}
  H_{\rm{HMHD}} & = &
  C_{C} \int
  \big[
    \HALL {\bm{V}} \cdot {\bm{\Omega}} 
    + 2 {\bm{V}} \cdot {\bm{B}}
  \big] {\rm{d}}^3\vec{x}
\nonumber\\&&
  +
  \frac{C_{M}}{\HALL}
  \int
  {\bm{A}} \cdot {\bm{B}} \,{\rm{d}}^3\vec{x}
.
\label{HMHD:derived helicity}
\end{eqnarray}
The constant $H$ becomes the magnetic helicity when 
$(C_{C},C_{M})=(0,\HALL)$,
while
the parameter value $(C_{C},C_{M})=(\HALL,\HALL)$ 
yields the hybrid helicity.\cite{turner1986hall}

\subsection{Section summary}

Therfore, the variation of the integral path 
due to the generalized vorticity 
preserves the value of the action $S_{\epsilon}=S_{0}.$
Noticing that the generalized velocity, momentum, vorticity, and the operators ${\widehat{M}}$ and ${\widehat{D}}$ are related as
$$
  {\vec{\bm{V}}}
  \mathop{\longrightarrow}^{\widehat{M}}
  {\U{\vec{\bm{M}}}}
  \mathop{\longrightarrow}^{\widehat{D}}
  {\vec{\bm{\Omega}}}
,
$$
it is useful to define the integro-differential operator given by the product $\widehat{W}:=\widehat{D}\widehat{M}$, 
which maps the generalized velocity 
to the generalized vorticity: 
$
  {\vec{\bm{\Omega}}}=\widehat{W}{\vec{\bm{V}}}
.
$

\begin{figure}
\centering
\includegraphics[width=20em]{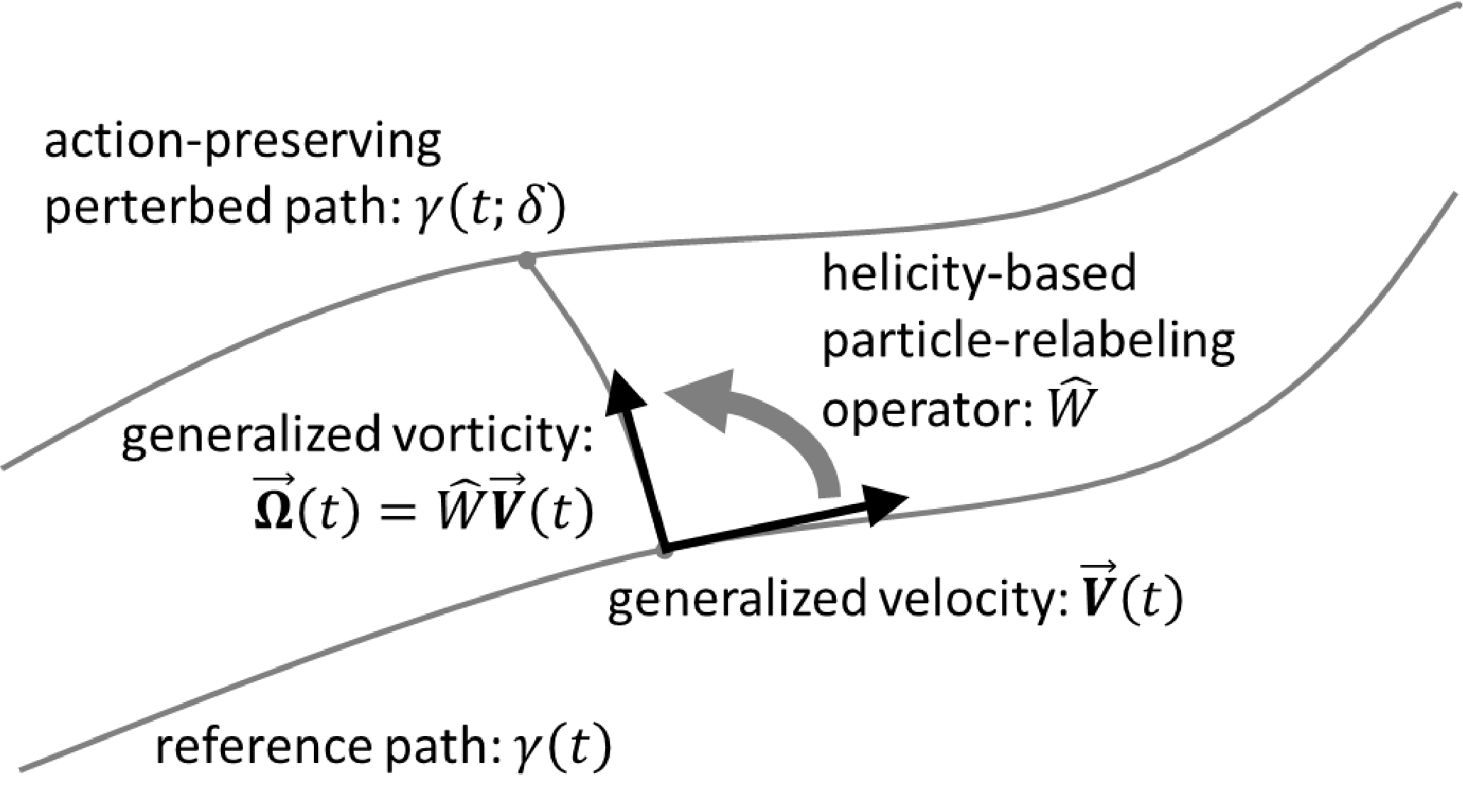}
\caption{\label{HMHD:fig}Relationship between the generalized velocities,
vorticities, and the reference and perturbed paths.}
\end{figure}
%
In Fig. \ref{HMHD:fig}, the relationship between the generalized velocities,
vorticities, and the reference and perturbed paths are presented.
Importantly, that the operator $\widehat{W}$ 
generates a field that satisfies the relabeling symmetry from the tangent vector of the reference path 
if the reference path locally satisfies the Euler-Lagrange equation.
Thus, we call the operator the \textit{helicity-based particle-relabeling operator}.

It is straightforward to check that the relabeling operator, $\widehat{W}$, is self-adjoint,
$
  \Braket{\big}{ \vec{\bm{V}}_{1} }{ \widehat{W}\vec{\bm{V}}_{2} }
  =
  \Braket{\big}{ \widehat{W}\vec{\bm{V}}_{1} }{ \vec{\bm{V}}_{2} }
,
$
for the HD, MHD, and HMHD systems.

\section{Eigenfunction of the helicity-based particle-relabeling operator and decomposition of the structure constant}

In this section, we show that the eigenequation of the helicity-based particle-relabeling operator leads to a common decomposition formula,
\begin{eqnarray}
  \Braket{\big}{
    {\vec{\bm{V}}}_{1}
  }{
    \LieBracket{\big}{{\vec{\bm{V}}}_{2}}{{\vec{\bm{V}}}_{3}}
  }
  =
  \Lambda
  \tripleHMHD{}{\vec{\bm{V}}_{1}}{\vec{\bm{V}}_{2}}{\vec{\bm{V}}_{3}}
,
\label{decomposition formula}
\end{eqnarray}
if ${\vec{\bm{V}}_{1}}$ is an eigenfunction of $\widehat{W}$,
where $\Lambda$ and the symbol 
$
  \tripleHMHD{}{\vec{\bm{V}}_{1}}{\vec{\bm{V}}_{2}}{\vec{\bm{V}}_{3}}
$
are the eigenvalue of the operator and the totally antisymmetric triple product of three ${\vec{\bm{V}}}$-variables, respectively.

Therefore, it is expected that the expressions of the Riemannian metric
and the structure constants are greatly simplified
if the variables are expanded by the eigenfunctions of $\widehat{W}$.

\subsection{Derivation of totally antisymmetric triple products
\label{Derivation of totally antisymmetric triple product}}
Integrating the combination of the Riemannian metrics and the Lie brackets by parts, we obtain the following expressions for each system:
\begin{eqnarray}
  \mbox{HD:}\ 
  \Braket{\big}{
    \bm{V}_{1}
  }{
    \LieBracket{\big}{\bm{V}_{2}}{\bm{V}_{3}}
  }
  &=&
  \int
    {\bm{\Omega}}_{1}
    \cdot({\bm{V}}_{2}\times{\bm{V}}_{3})
  {\rm{d}}^3\vec{x}
;
\label{HD:riemannMetricLieBracket}
\\
  \mbox{MHD:}\ 
  \Braket{\big}{
    {\vec{\bm{V}}}_{1}
  }{
    \LieBracket{\big}{{\vec{\bm{V}}}_{2}}{{\vec{\bm{V}}}_{3}}
  }
  &=&
  \int
  {\rm{d}}^3\vec{x} \Big[
    {\bm{\Omega}}_{1}
    \cdot({\bm{V}}_{2}\times{\bm{V}}_{3})
    +
    {\bm{B}}_{1}
   \nonumber\\&&
    \cdot \big(
      {\bm{V}}_{2} \times {\bm{J}}_{3} +
      {\bm{J}}_{2} \times {\bm{V}}_{3}
    \big)
  \Big]
;
\hspace{1em}
\label{MHD:riemannMetricLieBracket}
\\
  \mbox{HMHD:}\ 
  \Braket{\big}{
    {\vec{\bm{V}}}_{1}
  }{
    \LieBracket{\big}{{\vec{\bm{V}}}_{2}}{{\vec{\bm{V}}}_{3}}
  }
  &=&
  \frac{1}{\HALL}
  \int
  \Big\{
    ( \HALL {\bm{\Omega}}_{1} + {\bm{B}}_{1} )
    \cdot({\bm{V}}_{2}
   \nonumber\\&&\hspace{-11em}
    \times{\bm{V}}_{3})
    -
    {\bm{B}}_{1} \cdot \big[
      ({\bm{V}}_{2}-\HALL{\bm{J}}_{2}) \times
      ({\bm{V}}_{3}-\HALL{\bm{J}}_{3})
    \big]
  \Big\}
  {\rm{d}}^3\vec{x}
.
\label{HMHD:riemannMetricLieBracket}
\end{eqnarray}
The helicity-based particle-relabeling operators and the corresponding equations for the generalized velocities and vorticities are given by
\begin{eqnarray}
&&
  \widehat{W} = \curl
,\ 
\label{HD:W}
\\&&
  {\bm{\Omega}} = \Lambda {\bm{V}}
\label{HD:eigenequations1a}
\end{eqnarray}
for the HD system, 
\begin{eqnarray}
&&
  \widehat{W}
  =
  \left(\begin{array}{cc}
    O & - C_{C} (\HALL\curl)^{-1}
  \\
    -C_{C} \HALL\curl & C_{M} (\HALL\curl)^{-1}
  \end{array}\right)
,
\label{MHD:W}
\\&&
  {\bm{\Omega}} = -\frac{\tilde{C}_{M}\Lambda}{{C_C}\HALL}{\bm{V}} + \frac{\Lambda}{C_{C}} \bm{J},\ 
  {\bm{B}} = \frac{\Lambda}{C_{C}} {\bm{V}}
\label{MHD:eigenequations1a}
\end{eqnarray}
for the MHD system, and
\begin{eqnarray}
&&
  \widehat{W}
  =
  \left(\begin{array}{cc}
    C_{C} \HALL \curl & - C_{C} (\HALL \curl)^{-1}
  \\
    - C_{C} \HALL \curl & C_{M} (\HALL \curl)^{-1}
  \end{array}\right)
,
\label{HMHD:W}
\\&&
  \HALL {\bm{\Omega}} + {\bm{B}}
  =
  \frac{\Lambda}{C_{C}} \bm{V}
,\ 
  {\bm{B}}
  =
  \frac{\Lambda}{ C_{C} ( 1 - \tilde{C}_{M} ) }( {\bm{V}} - \HALL \bm{J} )
\label{HMHD:eigenequations1a}
\end{eqnarray}
for the HMHD system, 
{where $\tilde{C}_{M}:={C}_{M}/{C}_{C}$.}

Two remarks should be made. First, the solution of the eigenequation, Eq. (\ref{HMHD:eigenequations1a}), which agrees with Eq. (10) of Ref. \onlinecite{MahajanYoshida1998}, is given by the double Beltrami flow (DBF).
The DBF was introduced to describe force-free solutions of the HMHD system.
In our formulation, the force-free condition is easily verified, i.e., when an eigenfunction of $\widehat{W}$ is substituted into the quadratic term of the vorticity equation, it becomes zero:
$$
  \LieBracket{\big}{{\vec{\bm{\Omega}}}}{{\vec{\bm{V}}}}
  =
  \LieBracket{\big}{\widehat{W}{\vec{\bm{V}}}}{{\vec{\bm{V}}}}
  =
  \Lambda
  \LieBracket{\big}{{\vec{\bm{V}}}}{{\vec{\bm{V}}}}
  =
  {{\vec{\bm{0}}}}
.
$$

Second, the DBF has corresponding plasma states in the MHD system.
That is, if $\tilde{C}_{M}\approx O(\HALL)$, for sufficiently small $\HALL$, Eq. (\ref{HMHD:eigenequations1a}) reads as
\begin{eqnarray}
  \HALL {\bm{\Omega}}
  &=&
  -
  \frac{\Lambda}{ C_{C} ( 1 - \tilde{C}_{M} ) }( {\bm{V}} - \HALL \bm{J} )
  +
  \frac{\Lambda}{C_{C}} \bm{V}
\nonumber\\
  &\approx&
  -
  \frac{\tilde{C}_{M}\Lambda}{ C_{C} } {\bm{V}}
  +
  \frac{ \HALL \Lambda }{ C_{C} } \bm{J}
  +
  o(\HALL)
,
\nonumber\\
  {\bm{B}}
  &=&
  \frac{\Lambda}{ C_{C} ( 1 - \tilde{C}_{M} ) }( {\bm{V}} - \HALL \bm{J} )
  \approx
  \frac{\Lambda}{C_{C}}{\bm{V}}
  +
  O(\HALL)
,
\nonumber
\end{eqnarray}
i.e., the HMHD eigenmodes asymptote to those of the MHD system (Eq. (\ref{MHD:eigenequations1a})) in this limit, even though the corresponding helicity-based particle-relabeling operator $\widehat{W}$ does not.

Substituting these relations into Eqs.
(\ref{HD:riemannMetricLieBracket}), 
(\ref{MHD:riemannMetricLieBracket}), and
(\ref{HMHD:riemannMetricLieBracket}),
we verify that the decomposition formula, Eq. 
(\ref{decomposition formula}), 
holds for each system, where
\begin{eqnarray}
  \Lambda
  \tripleHMHD{}{\vec{\bm{V}}_{1}}{\vec{\bm{V}}_{2}}{\vec{\bm{V}}_{3}}
  =
  \Lambda
  \int
    {\bm{V}}_{1}
    \cdot({\bm{V}}_{2}\times{\bm{V}}_{3})
  {\rm{d}}^3\vec{x}
\label{HD:tripleProduct}
\end{eqnarray}
for the HD system,
\begin{eqnarray}
  \Lambda
  \tripleHMHD{}{\vec{\bm{V}}_{1}}{\vec{\bm{V}}_{2}}{\vec{\bm{V}}_{3}}
  &=&
  \frac{\Lambda
  }{C_{C}}
  \int
  \Big\{
    -
    \frac{\tilde{C}_{M}}{\HALL}
    {\bm{V}}_{1}\cdot({\bm{V}}_{2}\times{\bm{V}}_{3})
   \nonumber\\&&
    +
    \big[
      {\bm{J}}_{1}
      \cdot ( {\bm{V}}_{2} \times {\bm{V}}_{3})
      +
      {\bm{V}}_{1} \cdot ( {\bm{J}}_{2} \times {\bm{V}}_{3})
     \nonumber\\&&
      +
      {\bm{V}}_{1} \cdot ( {\bm{V}}_{2} \times {\bm{J}}_{3})
    \big]
  \Big\}
  {\rm{d}}^3\vec{x}
\end{eqnarray}
for the MHD system, and
\begin{eqnarray}
  \Lambda
  \tripleHMHD{}{\vec{\bm{V}}_{1}}{\vec{\bm{V}}_{2}}{\vec{\bm{V}}_{3}}
  &=&
  \frac{\Lambda
  }{C_{C}\HALL}
  \int
  \Big\{
    {\bm{V}}_{1}\cdot({\bm{V}}_{2}\times{\bm{V}}_{3})
    -
    \frac{1}{1-\tilde{C}_{M}}
   \nonumber\\&&\times
    ({\bm{V}}_{1}-\HALL{\bm{J}}_{1})
    \cdot 
    \big[
      ({\bm{V}}_{2}-\HALL{\bm{J}}_{2})
     \nonumber\\&&
      \times
      ({\bm{V}}_{3}-\HALL{\bm{J}}_{3})
    \big]
  \Big\}
  {\rm{d}}^3\vec{x}
\label{HMHD:tripleProduct}
\hspace{2em}
\end{eqnarray}
for the HMHD system.
Note that, if $\tilde{C}_{M}\approx O(\HALL)$, 
the asymptotic relation
\begin{eqnarray}
  \tripleHMHD{}{\vec{\bm{V}}_{1}}{\vec{\bm{V}}_{2}}{\vec{\bm{V}}_{3}
  }_{\rm{HMHD}}
  &\approx&
  \tripleHMHD{}{\vec{\bm{V}}_{1}}{\vec{\bm{V}}_{2}}{\vec{\bm{V}}_{3}
  }_{\rm{MHD}}
\label{asymp triple prod}
\end{eqnarray}
holds in the limit of $\HALL\to0$.

\subsection{Normal-mode representation of the Riemannian metric and triple product of the generalized velocities}

Let $\bm{\phi}=\bm{\phi}(\vec{x};\vec{k},\HEL)$ be a Beltrami flow (BF), i.e., an eigenfunction of the curl operator:
\begin{equation}
  \curl\bm{\phi}(\vec{x};\vec{k},\HEL)=\HEL\lambda\bm{\phi}(\vec{x};\vec{k},\HEL)
,
\label{curl eigenequation}
\end{equation}
where the symbols
$\vec{k}$, $\HEL=\pm1$, and $\lambda$ stand for the mode-identifier (e.g., the wavenumber vector for periodic boundary conditions or Euclidean space) and the sign and modulus of the eigenvalue corresponding to the mode $\vec{k}$, respectively.
Complex helical waves on a periodic box \cite{waleffe1992nature}
and Chandrasekhar-Kendall functions in a cylindrical configuration \cite{ChandrasekharKendall1957}
are well-known examples of this.
Note that, because BF $\bm{\phi}$ is solenoidal, it satisfies the identity $\curl\curl\bm{\phi}=-\triangle\bm{\phi}$ and, therefore, the modulus of the eigenvalue of the Laplacian, which is known to have negative eigenvalues, is expressed as $-\lambda^2$.
This implies that $\lambda$ is always a real number.

Because the helicity-based particle-relabeling operators $\widehat{W}$ contain the curl operator and its inverse, the eigenfunctions of $\widehat{W}$s
are described in terms of BF.
Substituting Eq. (\ref{curl eigenequation}) into the expressions of the operators, Eqs. (\ref{HD:W}), (\ref{MHD:W}), and (\ref{HMHD:W}), they are reduced to the number matrices
\begin{eqnarray}
&&
  \widehat{W} = \HEL\lambda
\end{eqnarray}
for the HD system, 
\begin{eqnarray}
&&
  \widehat{W}
  =
  \left(\begin{array}{cc}
    O & - C_{C} / \HALL\HEL\lambda
  \\
    -C_{C} \HALL\HEL\lambda & C_{M} / \HALL\HEL\lambda
  \end{array}\right)
\end{eqnarray}
for the MHD system, and
\begin{eqnarray}
&&
  \widehat{W}
  =
  \left(\begin{array}{cc}
    C_{C} \HALL\HEL\lambda & - C_{C} / \HALL\HEL\lambda
  \\
    - C_{C} \HALL\HEL\lambda & C_{M} / \HALL\HEL\lambda
  \end{array}\right)
\end{eqnarray}
for the HMHD system.
Their corresponding eigenvalues are all real valued and given by
\begin{eqnarray}
&&
  \Lambda = \HEL \lambda
,\ 
\end{eqnarray}
for the HD system, 
\begin{eqnarray}
&&
  \Lambda=
  \HEL\POL C_C \Bigg[
    \sqrt{
      1
      +
      \Big(\frac{\tilde{C}_{M}}{2\HALL\lambda}\Big)^2
    }
    +
    \frac{\POL \tilde{C}_{M}}{2\HALL\lambda}
  \Bigg]
,
\end{eqnarray}
for the MHD system, hereafter $\POL=\pm1$,
\begin{eqnarray}
  \Lambda=
  \HEL\POL C_C \Bigg[
    \sqrt{
      1
      + \Big(
        \frac{\HALL\lambda}{2}-\frac{\tilde{C}_{M}}{2\HALL\lambda}
      \Big)^2
    }
    +
    \POL
    \Big(\frac{\HALL\lambda}{2}+\frac{\tilde{C}_{M}}{2\HALL\lambda}\Big)
  \Bigg]
,
\nonumber\\
\label{Lambda MHD}
\end{eqnarray}
for the HMHD system, respectively.
If $\tilde{C}_{M}\approx O(\HALL)$, an asymptotic relation 
\begin{eqnarray}
  {\Lambda}_{\rm{HMHD}} \approx {\Lambda}_{\rm{MHD}}
\label{asymp triple lambda}
\end{eqnarray}
holds in the limit of $\HALL\to0$.
\newcommand{\LAM}[1]{\Lambda(#1)}
\newcommand{\ULAM}[1]{\underline{\Lambda}(#1)}
\newcommand{\KK}[1]{K(#1)}
We use the following eigenfunctions as the normal-modes:
\begin{eqnarray}
  \vec{\bm{\Phi}}_{a}
  &=&
  {\bm{\phi}}(\vec{x};\vec{k}_{a},\HEL_{a})
\end{eqnarray}
for the HD system and
\begin{eqnarray}
  \vec{\bm{\Phi}}_{a}
  &=&
  \vec{\bm{\Phi}}(\vec{x};\vec{k}_{a},\HEL_{a},\POL_{a})
\nonumber\\
  &:=&
  \left(\begin{array}{r}
    (\LAM{a}-{C_{M}}/{\KK{a}}) \bm{\phi}(\vec{x};\vec{k}_{a},\HEL_{a})
  \\
    -{C_{C}}{\KK{a}} \bm{\phi}(\vec{x};\vec{k}_{a},\HEL_{a})
  \end{array}\right)
\label{MHD,HMHD:eigenfunction CK CM}
\end{eqnarray}
for the MHD and HMHD systems,
where
$$
  \LAM{a}:=\Lambda(\vec{k}_{a},\HEL_{a},\POL_{a})
,\ \ 
  \KK{a} := \HALL \HEL_{a} \lambda_{a}
$$
are the eigenvalues of the operators $\widehat{W}$ and $\HALL\curl$, respectively, for the mode ``$a$''.
Hereafter, subscripts or superscripts of functions and coefficients abbreviate the mode indices.

The normal-mode-expansion of the generalized velocity is given by
\begin{eqnarray}
  {\vec{\bm{V}}}(\vec{x},t)
  =
  V^{\alpha}(t)\, {\vec{\bm{\Phi}}}_{\alpha}(\vec{x})
.
\end{eqnarray}
Hereafter, Einstein's summation convention is used for repeated Greek-letter subscript-superscript index pairs.
The values of each component of the Riemannian metric tensor and the structure constant of the Lie group are defined by
\begin{eqnarray}
  \M_{ab}
  & = &
  \Braket{\big}{{\vec{\bm{\Phi}}}_{a}}{{\vec{\bm{\Phi}}}_{b}}
,
\label{riemannianMetricPhi}
\\
  \LieBracket{\big}{{\vec{\bm{\Phi}}}_{b}}{{\vec{\bm{\Phi}}}_{c}}
  & = &
  C^{\alpha}_{bc}{\vec{\bm{\Phi}}}_{\alpha}
,
\label{lieBracketPhi}
\end{eqnarray}
where $a$, $b$, and $c$ stand for the mode indices of the eigenfunctions.
Substituting Eqs. (\ref{riemannianMetricPhi}) and (\ref{lieBracketPhi}) into Eq. (\ref{decomposition formula}), we obtain the following remarkable formula relating the Riemannian metric, the Lie bracket, and the eigenvalue of the relabeling operator:
\begin{eqnarray}
  \M_{a\beta}C^{\beta}_{bc}=\LAM{a}T_{abc}
,
\label{MC=LT}
\end{eqnarray}
where the three-index symbol $T_{abc}$ is the triple product of the eigenfunctions derived in the previous subsection:
\begin{eqnarray}
  T_{abc}
  & := &
  \tripleHMHD{\big}{{\vec{\bm{\Phi}}}_{a}}{{\vec{\bm{\Phi}}}_{b}}{{\vec{\bm{\Phi}}}_{c}}
.
\end{eqnarray}
Because the eigenfuctions of the curl operator are orthogonal to each other, the Riemannian metric and the triple product are given by
\begin{eqnarray}
  \M_{ab}^{\rm{HD}}
  & := &
  \int
    \bm{\phi}(\vec{k}_{a},\HEL_{a})
    \cdot
    \bm{\phi}(\vec{k}_{b},\HEL_{b})
  \,{\rm{d}}^3\vec{x}
  =
  \delta_{ab}
,
\\
  T_{abc}^{\rm{HD}}
  &=&
  \int
    \bm{\phi}(\vec{k}_{a},\HEL_{a})
    \cdot \big(
      \bm{\phi}(\vec{k}_{b},\HEL_{b}) \times
      \bm{\phi}(\vec{k}_{c},\HEL_{c})
    \big)
  {\rm{d}}^3\vec{x}
\hspace{1em}
\end{eqnarray}
for the HD system, 
\begin{eqnarray}
  \M_{ab}^{\rm{MHD}}
  & = &
  \Bigg[ \bigg( \LAM{a}-\frac{C_{M}}{\KK{a}} \bigg)^2 + C_{C}^2 \Bigg]\delta_{ab}
,
\\
  T_{abc}^{\rm{MHD}}
  & = &
  C_{C}^{-2}{\HALL}^{-1}
  T_{abc}^{\rm{HD}}\ULAM{a}\ULAM{b}\ULAM{c}
  \big(
    {C_{M}}
   \nonumber\\&&
    - \LAM{a}\KK{a}
    - \LAM{b}\KK{b}
    - \LAM{c}\KK{c}
  \big)
,
\label{MHD: T in Lambda}
\end{eqnarray}
for the MHD system, and
\begin{eqnarray}
  \M_{ab}^{\rm{HMHD}}
  & = &
  \Bigg[ \bigg( \LAM{a}-\frac{C_{M}}{\KK{a}} \bigg)^2 + C_{C}^2 \Bigg]\delta_{ab}
,
\\
  T_{abc}^{\rm{HMHD}}
  & = &
  \frac{T_{abc}^{\rm{HD}}}{\HALL C_{C}}
  \Bigg[
    \bigg( \LAM{a}-\frac{C_{M}}{\KK{a}} \bigg)
    \bigg( \LAM{b}-\frac{C_{M}}{\KK{b}} \bigg)
   \nonumber\\&&\times
    \bigg(
      \LAM{c}
      -\frac{C_{M}}{\KK{c}}
    \bigg)
    +
    \frac{\ULAM{a}\ULAM{b}\ULAM{c}}{1-\tilde{C}_{M}}
  \Bigg]
,
\label{HMHD: T in Lambda}
\end{eqnarray}
for the HMHD system, respectively.
Hereafter, the symbol $\U{\Lambda}$ in Eqs. 
(\ref{MHD: T in Lambda}) and (\ref{HMHD: T in Lambda})
denotes 
$$\ULAM{a}:=\Lambda(\vec{k}_{a},\HEL_{a},-\POL_{a}),$$
which is one of the two eigenvalues of the operator $\widehat{W}$ for assigned $\vec{k}_a$ and $\HEL_a$ other than $\Lambda(a)=\Lambda(\vec{k}_{a},\HEL_{a},\POL_{a}).$

\subsection{Normal-mode expansion coefficient representation of the evolution equations}
The coefficient of the generalized momentum is defined by the functional derivative of the Lagrangian with respect to the mode ${\vec{\bm{\Phi}}}_{a}$:
\begin{eqnarray}
  M_{a}(t)
  &:=&
  \left.
  \frac12\frac{\rm{d}}{{\rm{d}}\epsilon}
  \Braket{\big}{
    \vec{\bm{V}}(t) + \epsilon\vec{\bm{\Phi}}_{a}
  }{
    \vec{\bm{V}}(t) + \epsilon\vec{\bm{\Phi}}_{a}
  }
  \right|_{\epsilon=0}
\nonumber\\
  &=&
  \Braket{\big}{{\vec{\bm{\Phi}}}_{a}}{{\vec{\bm{V}}}(t)}
  =
  \Braket{\big}{{\vec{\bm{\Phi}}}_{a}}{V^{\beta}(t)\vec{\bm{\Phi}}_{\beta}}
  =
  \M_{a\beta}V^{\beta}(t)
.
\hspace{2em}
\end{eqnarray}
This equation also implies that the expansion coefficient $V^{a}(t)$ 
is obtained from
$
  V^{a}(t)=\M^{a\beta}\Braket{\big}{{\vec{\bm{\Phi}}}_{\beta}}{{\vec{\bm{V}}}(t)}
,
$
where $\M^{ab}$ is the inverse of $\M_{bc}$: 
$\M^{a\beta} \M_{\beta c}:=\delta^{a}_{c}.$

Using the inertia operator $\widehat{M}$, 
we introduce the base functions of the generalized momentum space, 
$\U{\vec{\bm{\Phi}}}{}^{a}$, each of which are defined by
\begin{eqnarray}
  \U{\vec{\bm{\Phi}}}{}^{a}
  &=&
  g^{a\beta} \widehat{M} \vec{\bm{\Phi}}_{\beta}
.
\end{eqnarray}
This definition comes from the following mode expansion of the generalized momentum:
$$
  \U{\vec{\bm{M}}}(t)
  = \widehat{M}\vec{\bm{V}}(t)
  = \widehat{M}(V^{\beta}(t) {\vec{\bm{\Phi}}}{}_{\beta})
  = V^{\beta} \M_{\beta\alpha} \U{\vec{\bm{\Phi}}}{}^{\alpha}
  = M_{\alpha}(t) \U{\vec{\bm{\Phi}}}{}^{\alpha}
.
$$
Each of the base functions, $\U{\vec{\bm{\Phi}}}{}^{a}$, is given by
\begin{eqnarray}
  \U{\vec{\bm{\Phi}}}{}^{a}
  &=&
  \bm{\phi}(\vec{x};\vec{k}_{a},\HEL_{a})
\end{eqnarray}
for the HD system and
\begin{eqnarray}
  \U{\vec{\bm{\Phi}}}{}^{a}
  &=&
  \frac{1}{g(a)}
  {\Big.}^t\!\!
  \left(\begin{array}{r}
    (\LAM{a}-{C_{M}}/{\KK{a}}) \bm{\phi}(\vec{x};\vec{k}_{a},\HEL_{a})
  \\
    - {C_{C}}/{\KK{a}}\, \bm{\phi}(\vec{x};\vec{k}_{a},\HEL_{a})
  \end{array}\right)
\end{eqnarray}
for the MHD and HMHD systems, where $g(a)$ is the diagonal element of $\widehat{M}$, i.e., $g_{ab}=g(a)\delta_{ab}$.

Using the bases $\{\vec{\bm{\Phi}}{}_{a}\}$ and $\{\U{\vec{\bm{\Phi}}}{}^{a}\}$,
the eigenvalue problem for $\widehat{W}$ 
can be rewritten as follows:
$$
  \widehat{W} {\vec{\bm{\Phi}}}{}_{a}
  = \widehat{D} \widehat{M} {\vec{\bm{\Phi}}}{}_{a}
  = \M_{a\beta} D^{\beta\gamma} \vec{\bm{\Phi}}_{\gamma}
  = \LAM{a} {\vec{\bm{\Phi}}}{}_{a}
,
$$
and therefore the matrix elements of the operators $\widehat{D}$ and $\widehat{W}$
are obtained as follows:
\begin{eqnarray}
  D^{ab} & = & \LAM{a} \M^{ab} = (\LAM{a}/g(a)) \delta^{ab}
,
\label{matrix element of D}
\\
  W^{a}_{b} & = & \LAM{a} \delta^{a}_{b}
.
\end{eqnarray}
Equations (\ref{MC=LT}) and (\ref{matrix element of D}) 
yield a decomposition formula for the structure constant of the Lie group such that 
\begin{eqnarray}
  C^{a}_{bc}
  = \LAM{a} \M^{a\beta} T_{\beta bc}
  = D^{a\beta} T_{\beta bc}
.
\label{relationCDB}
\end{eqnarray}


Taking the inner product of the Euler-Lagrange equation, Eq. 
(\ref{the euler-poincare eq}),
with the mode $\vec{\bm{\Phi}}_{i}$
and using Eq. (\ref{def:ad dagger}),
we obtain
\begin{eqnarray}
  \Braket{\big}{\vec{\bm{\Phi}}_{i}}{\dot{\vec{\bm{V}}}(t)}
  -
  \Braket{\big}{
    {\vec{\bm{V}}(t)}
  }{
    \LieBracket{\big}{\vec{\bm{\Phi}}_{i}}{\vec{\bm{V}}(t)}
  }
  =0
,
\end{eqnarray}
which yields the following evolution equation for the normal-mode expansion coefficients:
\begin{eqnarray}
  \M_{i\alpha}\dot{V}^{\alpha}(t)
  -
  \M_{\beta\gamma}\, C^{\gamma}_{i\lambda}\, V^{\beta}(t)\, V^{\lambda}(t)\,
  =
  0
.
\label{EulerLagrangeTDM}
\end{eqnarray}
Using Eq. (\ref{MC=LT}), this equation is rewritten as
\begin{eqnarray}
  \M_{i\alpha}\dot{V}^{\alpha}(t)
  =
  \LAM\beta\, T_{\beta i\lambda}\, V^{\beta}(t)\, V^{\lambda}(t)
.
\label{EulerLagrangeGeneral}
\end{eqnarray}
Because $\M_{ab}$ and $T_{abc}$ are symmetric and totally antisymmetric, respectively, this equation results in the following equations:
\begin{eqnarray}
  \frac{d}{dt}(\M_{\iota\alpha}{V}^{\iota}(t){V}^{\alpha}(t))
  &=&
  \frac{d}{dt}(\M_{\iota\alpha}\LAM\iota{V}^{\iota}(t){V}^{\alpha}(t))
  =
  0
,
\nonumber
\end{eqnarray}
which are the energy and helicity conservation laws, respectively.

In terms of the generalized momentum, 
Eq. (\ref{EulerLagrangeTDM}) yields the Euler-Poincare equation:
\cite{marsden2013introduction}
\begin{eqnarray}
  \dot{M}_{i}(t)
  -
  C^{\gamma}_{i\lambda}\, M_{\gamma}(t)  \, V^{\lambda}(t)
  =
  0
.
\label{EulerPoincareTDM}
\end{eqnarray}
Operating $D^{ji}$ on both sides of this equation, introducing the notation $\Omega^{j}(t) := D^{j\iota} M_{\iota}(t)$,
and using the decomposition relation, Eq. (\ref{relationCDB}),
we obtain the expansion coefficient expression of the generalized vorticity equations (or the particle-relabeling relation): 
\begin{eqnarray}
  \dot{\Omega}^{j}(t)
  +
  C^{j}_{\alpha\lambda} \, \Omega^{\alpha}(t) \, V^{\lambda}(t)
  =
  0
.
\end{eqnarray}

\section{Inclusion of the uniform external field effect}

Our formulation of the triplet of generalized velocities, momenta, and vorticity, in which the evolution equation is also given by the vorticity equation 
\begin{eqnarray}
  \ddt{\vec{\bm{\Omega}}}
  +
  \LieBracket{\big}{\vec{\bm{\Omega}}}{\vec{\bm{V}}}
  =
  {\vec{\bm{0}}}
,
\label{generalized vorticity equation 0}
\end{eqnarray}
has an interesting natural extension to non-solenoidal function space.

In the field of plasma physics,
stationary uniform external magnetic fields ($\bm{B}_{0}$) 
and the Coriolis force effect ($2\bm{\Omega}_{0}$) 
have attracted the attention of researchers.
For example, Mahajan et al. investigated the combination of the Coriolis force and the Hall-term effect on the dynamo process under the existence of a uniform magnetic field.\cite{mahajan2005wavescoriolis}

It is interesting that these uniform fields physically agree with the variables of generalized vorticities.
Therefore, introducing stationary uniform generalized vorticities given by
$$
  \vec{\bm{\Omega}}_0
  =
  \left(\begin{array}{c}
    C_C \bm{B}_0 \\ - 2 C_C \HALL \bm{\Omega}_0 - C_M \bm{B}_0
  \end{array}\right)
$$
for the MHD system and
$$
  \vec{\bm{\Omega}}_0
  =
  \left(\begin{array}{c}
    C_C ( 2 \HALL \bm{\Omega}_0 + \bm{B}_0 )
   \\
    - 2 C_C \HALL \bm{\Omega}_0 - C_M \bm{B}_0
  \end{array}\right)
$$
for the HMHD system and substituting 
$\vec{\bm{\Omega}}_0+\epsilon\vec{\bm{\Omega}}$ 
into Eq. (\ref{generalized vorticity equation 0}), 
we obtain the evolution equation of MHD/HMHD plasmas 
\begin{eqnarray}
  \epsilon
  \ddt{\vec{\bm{\Omega}}}
  +
  \epsilon
  \LieBracket{\big}{\vec{\bm{\Omega}}_0}{\vec{\bm{V}}}
  +
  \epsilon^2
  \LieBracket{\big}{\vec{\bm{\Omega}}}{\vec{\bm{V}}}
  =
  {\vec{\bm{0}}}
,
\label{vorticity eq with uniform BGF}
\end{eqnarray}
where $\epsilon\vec{\bm{\Omega}}=\epsilon\widehat{W}\vec{\bm{V}}$.
Mathematically, the uniform generalized vorticities belong to the function space of \textit{harmonic functions}, whose divergence and curl vanish.

\subsection{HMHD: double Beltrami wave modes\cite{mahajan2005wavescoriolis}}
At order $O(\epsilon)$, Eq.
(\ref{vorticity eq with uniform BGF})
reduces to $\ddt\vec{\bm{\Omega}}=-[\vec{\bm{\Omega}}_0,\vec{\bm{V}}]$.
For the HMHD system, we obtain the following linear wave equations:
\begin{eqnarray}
  \partial_t
  ( C_{C} \HALL {\bm{\Omega}} + C_{C} {\bm{B}} )
  &=&
  ( 2 C_{C} \HALL {\bm{\Omega}}_0 + C_{C} {\bm{B}}_0 )\cdot\nabla {\bm{V}}
,
\nonumber\\
  \partial_t
  (-C_{C}\HALL{\bm{\Omega}}-C_{M}{\bm{B}})
  &=&
  (-2C_{C}\HALL{\bm{\Omega}}_0-C_{M}{\bm{B}}_0)
  \cdot \nabla 
  {\bm{V}}
 \label{HMHD:BGF linear eqs}\\&&
  +
  (C_{C}-C_{M}){\bm{B}}_0 \cdot \nabla (-\HALL{\bm{J}})
.
\hspace{2em}
\nonumber
\end{eqnarray}
Here, we seek the solution in the Euclidean space $E^3$ or under the periodic boundary conditions $\mathbb{T}^3$, i.e., linear waves that have a functional form given by
\newcommand{\rmi}{{\rm{i}}}
\begin{eqnarray}
  \left(\begin{array}{c}
    \bm{V}(\vec{k},\HEL,\omega;\vec{x},t)
  \\
    -\HALL\bm{J}(\vec{k},\HEL,\omega;\vec{x},t)
  \end{array}\right)
  &=&
  \left(\begin{array}{c}
    \hat{V}
  \\
    -\HALL\hat{J}
  \end{array}\right)
  \bm{\phi}(\vec{x};\vec{k},\HEL)
  e^{-2\pi\rmi\omega t}
.\hspace{1.5em}
\nonumber
\end{eqnarray}
The function $\bm{\phi}$ designates the complex helical waves defined by
\cite{waleffe1992nature,araki2015differential}
\begin{eqnarray}
  \bm{\phi}(\vec{x};\vec{k},\HEL)
  &:=&
  2^{-1/2}
  (\bm{e}_{\theta}(\vec{k})+{\rm{i}}\HEL\bm{e}_{\phi}(\vec{k}))
  e^{2\pi{\rm{i}}\vec{k}\cdot\vec{x}}
,
\nonumber
\end{eqnarray}
where the symbols $\vec{k}=(k_x,k_y,k_z)$, $\bm{e}_{\theta}$, and $\bm{e}_{\phi}$ are the wavenumber vector and the unit vectors of the $\theta$- and $\phi$-directions in wavenumber space, respectively.

\renewcommand{\sakujo}[1]{}
\sakujo{
\vs
Assuming that the uniform fields are directed in z-direction ($\bm{B}_0=B_0\hat{z}$, $\bm{\Omega}_0=\Omega_0\hat{z}$) and $C_C\ne0$, the linear wave equation (\ref{HMHD:BGF linear eqs}) is reduced to 
\newcommand{\Kk}{\HALL \HEL |\vec{k}|}
\begin{eqnarray}
&&
  -2\pi\rmi\omega
  \left(\begin{array}{cc}
    \Kk  & - 1 / \Kk \\ - \Kk & \tilde{C}_M / \Kk
  \end{array}\right)
  \left(\begin{array}{c} \hat{V} \\ -\HALL\hat{J} \end{array}\right)
 \nonumber\\&&
  =
  2\pi\rmi k_z
  \left(\begin{array}{cc}
    2 \HALL \Omega_0 + B_0 & O
  \\
    - 2 \HALL \Omega_0 - \tilde{C}_M B_0 & (1 - \tilde{C}_M) B_0
  \end{array}\right)
  \left(\begin{array}{c} \hat{V} \\ -\HALL\hat{J} \end{array}\right)
.
\nonumber
\end{eqnarray}
Furthermore, if $C_C\ne C_M$, multiplying the 2$\times$2 matrix
$$
  \frac{1}{1 - \tilde{C}_M}
  \left(\begin{array}{cc}
    (1 - \tilde{C}_M)B_0 & 0
  \\
    2 \HALL{\Omega}_0 + \tilde{C}_M B_0  & 2 \HALL \Omega_0 + B_0
  \end{array}\right)
$$
on both sides, this equation is simplified as
$$
  \left(
    \omega W_0^{{\rm{HMHD}}}
    +
    k_z B_0(2\HALL{\Omega}_0 + B_0) I
  \right)
  \left(\begin{array}{c} \hat{V} \\ -\HALL\hat{J} \end{array}\right)
  =
  \left(\begin{array}{c} 0 \\ 0 \end{array}\right)
,
$$
where
$$
  W_0^{{\rm{HMHD}}}=
    \left(\begin{array}{cc}
      B_0 \Kk & - B_0 / \Kk \\ - B_0 \Kk & - 2 \HALL \Omega_0 / \Kk
    \end{array}\right)
,
$$
irrespective of the values of $C_C$ and $C_M$.
\vs
}\renewcommand{\sakujo}[1]{}

\ifx\Kk\undefined\newcommand{\Kk}{}\fi
\renewcommand{\Kk}{\HALL \HEL |\vec{k}|}
Assuming that the uniform fields are directed in the z-direction ($\bm{B}_0=B_0\hat{z}$, $\bm{\Omega}_0=\Omega_0\hat{z}$), $C_C\ne0$, and $C_C\ne C_M$, the linear wave equation, Eq. (\ref{HMHD:BGF linear eqs}), is simplified as follows:
$$
  \left(
    \omega W_0^{{\rm{HMHD}}}
    +
    k_z B_0(2\HALL{\Omega}_0 + B_0) I
  \right)
  \left(\begin{array}{c} \hat{V} \\ -\HALL\hat{J} \end{array}\right)
  =
  \left(\begin{array}{c} 0 \\ 0 \end{array}\right)
,
$$
where
\begin{eqnarray}
  W_0^{{\rm{HMHD}}}=
    \left(\begin{array}{cc}
      2\pi B_0 \Kk & - B_0 / 2\pi \Kk
    \\
      - 2\pi B_0 \Kk & - \HALL \Omega_0 / \pi \Kk
    \end{array}\right)
,
\end{eqnarray}
irrespective of the values of $C_C$ and $C_M$.
It is interesting that the obtained matrix is also the helicity-based particle-relabeling operator with coefficients $C_C=B_0$ and $C_M=-2\HALL\Omega_0$.
Therefore, the eigenfunctions of the relabeling operator, Eq. (\ref{MHD,HMHD:eigenfunction CK CM}), give the functional form of the linear waves.

Setting the vector ${}^t\!( \hat{V}, -\HALL\hat{J} )$ to be the eigenvector of $W_0^{{\rm{HMHD}}}$ and noticing that ${\rm{det}}(W_0^{{\rm{HMHD}}})=-B_0(2\HALL\Omega_0+B_0)=\Lambda \U\Lambda$, we can reduce this equation to
\begin{eqnarray}
  \left(\omega \Lambda - k_z \Lambda \U\Lambda \right)
  \left(\begin{array}{c} \hat{V} \\ -\HALL\hat{J} \end{array}\right)
  =
  \left(\begin{array}{c} 0 \\ 0 \end{array}\right)
,
\label{HMHD:eigenmode wave equation}
\end{eqnarray}
which leads to an interesting consequence: 
\begin{eqnarray}
  \omega(\vec{k},\HEL,\POL)
  =
  k_z \U\Lambda(\vec{k},\HEL,\POL)
.
\nonumber
\end{eqnarray}
Therefore, the functional form of the linear wave for $(\vec{k},\HEL,\POL)$ is given by
\begin{eqnarray}
  \left(\begin{array}{c}
    \bm{V}(\vec{k},\HEL,\POL;\vec{x},t)
  \\
    -\HALL\bm{J}(\vec{k},\HEL,\POL;\vec{x},t)
  \end{array}\right)
  & \propto & 
  \left(\begin{array}{c}
    \Lambda(\vec{k},\HEL,\POL) + {\Omega_0} / {\pi\HEL|\vec{k}|}
  \\
    -2\pi\HALL B_0\HEL|\vec{k}|
  \end{array}\right)
 \nonumber\\&&\times
  \bm{\phi}(\vec{x};\vec{k},\HEL)
  e^{-2\pi\rmi k_z\U\Lambda(\vec{k},\HEL,\POL) t}
,
\hspace{2em}
\label{eigenfunction HMHD B Omega}
\end{eqnarray}
where the explicit expression of the wave frequency is 
\begin{eqnarray}
  \omega(\vec{k},\HEL,\POL)
  &=&
  k_z \U\Lambda(\vec{k},\HEL,\POL)
  =
  k_z \Lambda(\vec{k},\HEL,-\POL)
\nonumber\\
  &=&
  - \HEL \POL B_0 k_z \Bigg[
    \sqrt{
      1
      +
      \Big(
        \pi\HALL|\vec{k}| + \frac{\Omega_0}{2\pi B_0|\vec{k}|}
      \Big)^2
    }
   \nonumber\\&&
    -
    \POL\Big(
      \pi\HALL|\vec{k}| - \frac{\Omega_0}{2\pi B_0|\vec{k}|}
    \Big)
  \Bigg]
.
\end{eqnarray}
Note that the waves obtained are DBFs.

The expression of the phase velocity implies that the linear waves are divided into two classes according to the sign of $\POL$.
If the Coriolis force is absent, the eigenfrequency is reduced to 
$
  \omega=
  - \HEL\POL B_0 k_z \Big(
    \sqrt{ 1 + (\pi\HALL|\vec{k}|/2)^2 }
    -
    \POL \pi \HALL |\vec{k}|/2
  \Big)
$
and the linear modes given by Eq. (\ref{eigenfunction HMHD B Omega}) converge to the generalized Els\"asser variables.\cite{galtier2006wave}
When $s=+1$, the phase velocities of the waves are relatively slow and correspond to \textit{ion cyclotron waves}.
Conversely, when $s=-1$, the corresponding waves belong to the fast phase velocity branch and are called \textit{whistler waves}.

\subsection{MHD}
For the MHD system, the leading order linear equations are
\begin{eqnarray}
  C_{C} \partial_t
  {\bm{B}}
  &=&
  C_{C} {\bm{B}}_0 \cdot\nabla {\bm{V}}
,
\nonumber\\
  \partial_t
  (-C_{C}\HALL{\bm{\Omega}}-C_{M}{\bm{B}})
  &=&
  (-2C_{C}\HALL{\bm{\Omega}}_0-C_{M}{\bm{B}}_0)
  \cdot \nabla 
  {\bm{V}}\hspace{2em}
 \label{MHD:BGF linear eqs}\\&&
  +
  C_{C} {\bm{B}}_0 \cdot \nabla (-\HALL{\bm{J}})
.
\nonumber
\end{eqnarray}
\renewcommand{\sakujo}[1]{}%
\sakujo{
Applying the same calculation procedures and associated notations as those used in the HMHD system case, we obtain
\begin{eqnarray}
&&
  -\omega
  \left(\begin{array}{cc}
    0 & - 1 / 2\pi\Kk \\ - 2\pi\Kk & \tilde{C}_M / 2\pi\Kk
  \end{array}\right)
  \left(\begin{array}{c} \hat{V} \\ -\HALL\hat{J} \end{array}\right)
 \nonumber\\&&
  =
  k_z
  \left(\begin{array}{cc}
    B_0 & O
  \\
    - ( 2 \HALL \Omega_0 + \tilde{C}_M B_0 ) & B_0
  \end{array}\right)
  \left(\begin{array}{c} \hat{V} \\ -\HALL\hat{J} \end{array}\right)
,
\nonumber
\end{eqnarray}
}%
If $C_C\ne0$, $\bm{B}_0=B_0\hat{z}$, and $\bm{\Omega}_0=\Omega_0\hat{z}$,
this equation is simplified as follows:
$$
  \left(
    \omega W_0^{{\rm{MHD}}}
    +
    k_z B_0^2 I
  \right)
  \left(\begin{array}{c} \hat{V} \\ -\HALL\hat{J} \end{array}\right)
  =
  \left(\begin{array}{c} 0 \\ 0 \end{array}\right)
,
$$
where
\begin{eqnarray}
  W_0^{{\rm{MHD}}}=
    \left(\begin{array}{cc}
      0 & - B_0 / 2\pi\Kk \\ - 2\pi B_0 \Kk & - \HALL \Omega_0 / \pi\Kk
    \end{array}\right)
,
\end{eqnarray}
irrespective of the values of $C_C$ and $C_M$.
The relation ${\rm{det}}(W_0^{{\rm{MHD}}})=-B_0^2=\Lambda \U\Lambda$ 
leads to the equation
\begin{eqnarray}
  \left(\omega \Lambda - k_z \Lambda \U\Lambda \right)
  \left(\begin{array}{c} \hat{V} \\ -\HALL\hat{J} \end{array}\right)
  =
  \left(\begin{array}{c} 0 \\ 0 \end{array}\right)
,
\label{MHD:eigenmode wave equation}
\end{eqnarray}
if the vector is chosen to be the eigenvector of $W_0^{{\rm{MHD}}}$.
Accordingly, the appearance of the linear wave formally agrees with that in the HMHD system:
\begin{eqnarray}
  \left(\begin{array}{c}
    \bm{V}(\vec{k},\HEL,\POL;\vec{x},t)
  \\
    -\HALL\bm{J}(\vec{k},\HEL,\POL;\vec{x},t)
  \end{array}\right)
  & \propto & 
  \left(\begin{array}{c}
    \Lambda(\vec{k},\HEL,\POL) + {\Omega_0} / {\pi\HEL|\vec{k}|}
  \\
    -2\pi\HALL B_0\HEL|\vec{k}|
  \end{array}\right)
 \nonumber\\&&\times
  \bm{\phi}(\vec{x};\vec{k},\HEL)
  e^{-2\pi\rmi k_z\U\Lambda(\vec{k},\HEL,\POL) t}
,
\hspace{2em}
\label{eigenfunction MHD B Omega}
\end{eqnarray}
where the explicit expression of the phase velocity is 
\begin{eqnarray}
&&
  \omega(\vec{k},\HEL,\POL)
  =
  k_z \U\Lambda(\vec{k},\HEL,\POL)
  =
  k_z \Lambda(\vec{k},\HEL,-\POL)
\nonumber\\&&\hspace{2em}
  =
  - \HEL\POL B_0 k_z \Bigg[
    \sqrt{
      1
      +
      \Big(\frac{\Omega_0}{2\pi B_0|\vec{k}|}\Big)^2
    }
    -
    \frac{\POL\,\Omega_0}{2\pi B_0|\vec{k}|}
  \Bigg]
.\hspace{2em}
\end{eqnarray}
Therefore, the linear waves in the MHD system are also obtained as the $\HALL\to0$ limit of those in the HMHD system.
If the Coriolis force is absent, the two branches of the phase velocities degenerate to $\omega=-\HEL\POL B_0 k_z$, which corresponds to \textit{{\Alfven} waves}.

\subsection{Inclusion of the linear wave propagation in the evolution equation}

For the HMHD case, the functional form of the linear waves under the existence of uniform fields $\bm{B}_0={B}_0\hat{z}$ and $\bm{\Omega}_0={\Omega}_0\hat{z}$ is given by the DBF; whereas, for the MHD case, the functional form is given by their $\HALL\to0$ limit.
Substituting the generalized velocity expanded by the eigenfunctions of 
$W_0^{{\rm{HMHD}}}$ or $W_0^{{\rm{MHD}}}$ for $B_{0}\ne0$,
\begin{eqnarray}
&&
  \left(\begin{array}{c}
    \bm{V}(\vec{x},t)
  \\
    -\HALL\bm{J}(\vec{x},t)
  \end{array}\right)
  =
  \sum_{\vec{k},\HEL,\POL}
  V(t;\vec{k},\HEL,\POL)\,
  \vec{\bm{\Phi}}_{0}(\vec{x};\vec{k},\HEL,\POL)
,
\nonumber
\end{eqnarray}
where
\begin{eqnarray}
  \vec{\bm{\Phi}}_{0}(\vec{x};\vec{k},\HEL,\POL)
  =
  \left(\begin{array}{c}\displaystyle
    \Lambda(\vec{k},\HEL,\POL) + \frac{\Omega_0}{\pi\HEL|\vec{k}|}
  \\
    -2\pi\HALL B_0\HEL|\vec{k}|
  \end{array}\right)
  \bm{\phi}(\vec{x};\vec{k},\HEL)
,
\nonumber
\end{eqnarray}
into Eq. (\ref{vorticity eq with uniform BGF}), using the following eigenvalue problem relations,
\begin{eqnarray}
  \vec{\bm{\Omega}}(\vec{x},t)
  &=&
  \sum_{\vec{k},\HEL,\POL}
  \Lambda(\vec{k},\HEL,\POL)\,
  V(t;\vec{k},\HEL,\POL)\,
  \vec{\bm{\Phi}}_{0}(\vec{x};\vec{k},\HEL,\POL)
,
\nonumber
\\
  \big[\vec{\bm{\Omega}}_0,\vec{\bm{V}}\big]
  &=&
  -2\pi\rmi 
  \sum_{\vec{k},\HEL,\POL}
  k_z \Lambda(t;\vec{k},\HEL,\POL)\, \U\Lambda(t;\vec{k},\HEL,\POL)\,
 \nonumber\\&&\times
  V(t;\vec{k},\HEL,\POL)\,
  \vec{\bm{\Phi}}_{0}(\vec{x};\vec{k},\HEL,\POL)
,
\nonumber
\end{eqnarray}
and taking the inner product with $\vec{\bm{\Phi}}_{0}(\vec{k},\HEL_k,\POL_k)$, 
we obtain the normal-mode expansion of Eq. (\ref{vorticity eq with uniform BGF}) as follows:
\newcommand{\DBM}[1]{\vec{#1},\HEL_{#1},\POL_{#1}}
\newcommand{\DBMx}[3]{#2\vec{#1},\HEL_{#1},#3\POL_{#1}}
\begin{widetext}
\begin{eqnarray}
&&
  \Braket{\big}
    {\vec{\bm{\Phi}}_{0}(\vec{k},\HEL_{k},\POL_{k})}
    {\vec{\bm{\Phi}}_{0}(-\vec{k},\HEL_{k},\POL_{k})}\,
  \Lambda(-\vec{k},\HEL_{k},\POL_{k})\,
  \Big(
    \ddt
    -
    2 \pi \rmi k_z\, \Lambda(-\vec{k},\HEL_{k},-\POL_{k})
  \Big)
  V(t;-\vec{k},\HEL_{k},\POL_{k})\,
  \nonumber\\&&
  +\Lambda(\vec{k},\HEL_{k},\POL_{k})\,
  \mathop{
    \sum_{\DBM{p}}
    \sum_{\DBM{q}}
  }^{\vec{k}+\vec{p}+\vec{q}=\vec{0}}
  \tripleHMHD{\big}
  {\vec{\bm{\Phi}}_{0}(\vec{k},\HEL_{k},\POL_{k})}
  {\vec{\bm{\Phi}}_{0}(\vec{p},\HEL_{p},\POL_{p})}
  {\vec{\bm{\Phi}}_{0}(\vec{q},\HEL_{q},\POL_{q})}\,
  \Lambda(\vec{p},\HEL_{p},\POL_{p})\,
  V(t;\vec{p},\HEL_{p},\POL_{p})\,
  V(t;\vec{q},\HEL_{q},\POL_{q})\,
  =0
.
\hspace{2em}
\label{evolution equation}
\end{eqnarray}
\end{widetext}
This equation can be reduced to
\begin{eqnarray}
&&
  \ddt
  \overline{U(t;\DBM{k})}\,
  +
  \mathop{
    \sum_{\DBM{p}}
    \sum_{\DBM{q}}
  }^{\vec{k}+\vec{p}+\vec{q}=\vec{0}}
  e^{-2\pi\rmi\Psi t}
 \nonumber\\&&\times
  \frac{
    \tripleHMHD{\big}
    {\vec{\bm{\Phi}}_{0}(\DBM{k})}
    {\vec{\bm{\Phi}}_{0}(\DBM{p})}
    {\vec{\bm{\Phi}}_{0}(\DBM{q})}
  }{
    \Braket{\big}
    {\vec{\bm{\Phi}}_{0}(\DBM{k})}
    {\vec{\bm{\Phi}}_{0}(-\DBM{k})}
  }
 \nonumber\\&&\times
  \Lambda(\DBM{p})\,
  U(t;\DBM{p})\,
  U(t;\DBM{q})\,
  =0
,
\end{eqnarray}
where the overbar indicates the complex conjugate, the variable $U$ and the phase factor $\Psi$ are defined by
\begin{eqnarray}
&&
  U(t;\DBMx{k}{}{})
  :=
  V(t;\DBMx{k}{}{})
  e^{2\pi\rmi\,k_z \U\Lambda(\DBMx{k}{}{}) t}
,
\nonumber
\\&&
  \Psi
  :=
  k_z\,\U\Lambda(\DBMx{k}{}{})+
  p_z\,\U\Lambda(\DBMx{p}{}{})+
  q_z\,\U\Lambda(\DBMx{q}{}{})
,
\nonumber
\end{eqnarray}
respectively, and the following relations are used:
$$
  V(t;-\vec{k},\HEL,\POL)=\overline{V(t;\vec{k},\HEL,\POL)},
\ \ 
  \Lambda(-\vec{k},\HEL,\POL)=\Lambda(\vec{k},\HEL,\POL)
.
$$
The resonant conditions are
\begin{eqnarray}
&&
  \vec{k}+\vec{p}+\vec{q}=\vec{0}
,
\\&&
  k_z\,\U\Lambda(\DBMx{k}{}{})+
  p_z\,\U\Lambda(\DBMx{p}{}{})+
  q_z\,\U\Lambda(\DBMx{q}{}{})=0
.
\nonumber\\
\label{resonant phase condition}
\end{eqnarray}
Therefore, the effect of the external uniform fields, $\bm{B}_0$ and $\bm{\Omega}_0$,
is reduced to the general formula, Eq. (\ref{EulerLagrangeGeneral}),
with the additive resonant phase condition, Eq. (\ref{resonant phase condition}).

\section{Discussion}

\subsection{Extension of relabeling symmetry and its physical implications}

In the present study, using the ideal incompressible HD, MHD, and HMHD systems as models, we demonstrated that the particle relabeling operation in HD is extendable to MHD and HMHD.
\vs

Considering the Lin constraints $v=\dot\xi+[\xi,V]$ on the semidirect product group, we extended the ``relabeling symmetry'' beyond a mere change in the Lagrangian labels of the fluid parcels.
\vs

This relation mathematically implies that, for the dynamical systems in some non-Abelian Lie groups, the time derivative of a perturbation, $\dot\xi$, and the velocity perturbation, $v$, differ by the commutator between the perturbation and the reference velocity, $[\xi,V]$.
Therefore, perturbations $\xi(t)$ that do not alter the reference solution $V(t)$ are possible.
One of the physical implications is that, even though the Lagrangian specification is redundant, this is a gateway to the general dynamical structure of such a system.

A very interesting finding of this study is that the HD, MHD, and HMHD systems were shown, in a somewhat heuristic way, to have such an integro-differential operator $\widehat{W}$ that the field $\Omega:=\widehat{W}V$ satisfies the particle-relabeling symmetry, i.e., $\dot\Omega+[\Omega,V]=0$ if $V$ is a solution of the Euler-Lagrange equation, Eq. (\ref{the euler-poincare eq}).
This leads to the well-known conservation laws of helicities, including the magnetic and cross helicities for MHD and the magnetic and hybrid helicities for HMHD; our approach reveals that these helicities are understandable as a consequence of a general mathematical structure.

Their common mathematical structure is revealed via a normal-mode expansion of the physical quantities and equations.
Because the eigenfunctions of the relabeling operators $\widehat{W}$ constitute the orthogonal bases of the HD, MHD, and HMHD systems, they function as normal modes of these dynamical systems.
Normal-mode expansions of the quantities of the HD, MHD, and HMHD systems reveal that the products of the Riemannian metrics and the Lie brackets, which are the foundations that define these dynamical systems, have a common decomposition formula
\begin{eqnarray}
  \M_{a\beta}C^{\beta}_{cd}=\Lambda(a)T_{acd}
,
\nonumber
\end{eqnarray}
where $\Lambda(a)$ is the eigenvalue of $\widehat{W}$.
This structure connects the evolution equations of the generalized velocities (or the Euler-Poincar\'e equations) to those of the generalized vorticities; in addition, this simplifies the expressions of the energy and helicity:
\begin{eqnarray}
  E=\frac12g_{\alpha\beta}V^{\alpha}V^{\beta}
,\hspace{1em}
  H=g_{\alpha\beta}\Lambda(\alpha)V^{\alpha}V^{\beta}
,
\nonumber
\end{eqnarray}
where $V^{\alpha}$s are the expansion coefficients of the generalized velocity.

This also enables us to discuss the so-called triad-interaction in a unified, general expression as follows:
\begin{eqnarray}
  g_{kk}\dot V^k & = &  (\Lambda(q)-\Lambda(p)) T_{kpq} V^p V^q ,\nonumber\\
  g_{pp}\dot V^p & = &  (\Lambda(k)-\Lambda(q)) T_{kpq} V^q V^k ,\\
  g_{qq}\dot V^q & = &  (\Lambda(p)-\Lambda(k)) T_{kpq} V^k V^p ,\nonumber
\end{eqnarray}
where $k$, $p$, and $q$ are the mode identifiers (e.g., the set of wavenumber, helicity, and wave mode, respectively) that satisfy the resonant conditions.
These triad-truncated model equations are reduced to 
\begin{eqnarray}
  \partial_{t^*} U^k & = &  (\Lambda(q)-\Lambda(p)) U^p U^q ,\nonumber\\
  \partial_{t^*} U^p & = &  (\Lambda(k)-\Lambda(q)) U^q U^k ,
\label{triad general reduced}
\\
  \partial_{t^*} U^q & = &  (\Lambda(p)-\Lambda(k)) U^k U^p ,\nonumber
\end{eqnarray}
if the variables $V$ and $t$ are changed to
\begin{eqnarray}
  U^k := \sqrt{g_{kk}}\,V^k e^{i\psi(k,\{k,p,q\})}
,
&&\ \ 
  t^*:=\frac{\sqrt{g_{kk}}\sqrt{g_{pp}}\sqrt{g_{qq}}}{|T_{kpq}|}t
,
\nonumber
\end{eqnarray}
where $\psi(k,\{k,p,q\})$ is an appropriate phase factor that depends on the shape and configuration of the triad $\{k,p,q\}$.
Therefore, the totally antisymmetric tensor $T_{kpq}$ physically defines the basic characteristic time scale for the quadratic mode interaction of the triad $\{k,p,q\}$.

\subsection{Relabeling operators, equilibrium MHD/HMHD plasma configurations, and corresponding helicities}

Double Beltrami flows (DBFs) were well-known to have been derived as an equilibrium solution to the HMHD equation having a non-zero flow as well as a magnetic field.\cite{MahajanYoshida1998}
In our previous study, the DBF was given as an eigenfunction of the helicity-based particle-relabeling operator of the HMHD system,\cite{araki2015helicity} while in the present study, the DBF was shown to have an MHD counterpart in the $\HALL\to0$ limit (even though the operator itself did not).

The helicity-based particle-relabeling operators for MHD and HMHD plasmas and their associated generalized vorticities have two arbitrary constants, $C_C$ and $C_M$ (see Eqs. (\ref{MHD:generalizedVorticity}), (\ref{HMHD:generalized vorticity}), (\ref{MHD:W}), and (\ref{HMHD:W})).
What is the meaning of these constants?

One of the fundamental properties of the generalized vorticity is that it is an action-preserving perturbation in the configuration space.
Action-preserving group operations are given by 
$$
  (g,h)=
  \big(e^{\tau C_C{\bm{B}}},-\HALL\tau(-C_C \HALL {\bm{\Omega}} - C_M {\bm{B}})\big)
$$
for the MHD system and 
$$
  (g,h)=
  \big(e^{\tau C_C(\HALL {\bm{\Omega}}+{\bm{B}})},e^{\tau(-C_C \HALL {\bm{\Omega}} - C_M {\bm{B}})}\big)
$$
for the HMHD system, where $\tau$ is a small perturbation parameter
(cf. Eqs. (\ref{MHD:group action def}) and (\ref{HMHD:group action def})).
These expressions indicate that the constant $C_C$ is related to the action of the whole semidirect product group $G\ltimes H$, i.e., it corresponds to the perturbation of both the velocity and the current field, while the constant $C_M$ appears only in the second component ($H$ of $G\ltimes H$), i.e., it affects only the current field.
Therefore, the two arbitrary constants $C_C$ and $C_M$ correspond to two independent group actions $(e^{\epsilon\xi},\,e^{-\epsilon\HALL\eta})$ and $(id,\,e^{-\epsilon\HALL\eta})$.

This is a generalization of our previous result for the HMHD system that each of the two independent relabeling operations yields a corresponding helicity, i.e., the relabeling of the ion fluid yields the hybrid helicity conservation, while the relabeling of the electron fluid yields magnetic helicity conservation.\cite{araki2015helicity}
\sakujo{
\vs

When DBF is chosen as the generalized vorticity, its relabeling occurs along the reference solution path.
That is, the displacement of plasma fluid particle in the direction of the generalized vorticity describes the plasma motion of the reference solution itself.
Thus, the DBF for assigned $C_C$ and $C_M$ describes such the stationary plasma motion that retains the value of the combination of helicities given by Eq. (\ref{MHD:derived helicity}) for MHD or (\ref{HMHD:derived helicity}) for HMHD.
}

\subsection{The weak interaction conjecture of MHD turbulence}

To conclude this study, two remarks need to be made:
one involves the application of Eq. (\ref{triad general reduced}) to the MHD system and the other is the possibility of extending our result to other systems.
\vs

When the Coriolis force is absent ($\bm{\Omega}_0=\bm{0}$) and the {\Alfven} waves, which correspond to $C_C=B_0$ (or $C_C=1$ for $\bm{B}_0=\bm{0}$) and $C_M=0$ for Eqs. (\ref{MHD,HMHD:eigenfunction CK CM}) or (\ref{eigenfunction MHD B Omega}), are chosen as the expansion functions, their corresponding eigenvalues become $\Lambda=\pm1$ irrespective of the wavenumber.
This results in only two cases, all three eigenvalues coincide ($\Lambda(k)=\Lambda(p)=\Lambda(q)=\pm1$):
\begin{eqnarray}
  \dot U^k =  \dot U^p = \dot U^q = 0,
\nonumber
\end{eqnarray}
or two of the three eigenvalues are the same 
($-\Lambda(k)=\Lambda(p)=\Lambda(q)=\pm1$):
\begin{eqnarray}
  \dot U^k & = & 0 ,\nonumber\\
  \dot U^p & = & \pm(2 U^k) U^q ,\nonumber\\
  \dot U^q & = & \mp(2 U^k) U^p ,\nonumber
\end{eqnarray}
which generates only periodic motions with constant angular frequency $\pm2U^k$.
This suggests that the mode interaction of the MHD system is somewhat ``weaker'' than that of the HD and HMHD systems, in which the three eigenvalues take different values in general.
In other words, the famous investigation by Waleffe\cite{waleffe1992nature} concerning energy tranfer in a fully developed turbulence via the nonlinear term does not seem to be the case for the MHD system.
Energy is transferred between only two of three modes at the level of the triad-truncated model.

\subsection{A note on the possibility of an extension to other systems}

Are there other examples of this type of dynamical system?
Here, we pick two well-known cases: the freely rotating top described by the Euler equation and the KdV equation formulated based on Bott-Virasoro algebra.

The Euler equation for a freely rotating top is known as a dynamical system of $SO(3)$, whose Lie bracket and the inner product are given by
\begin{eqnarray}
&&
  \LieBracket{\big}{u}{v}^i
  =\delta^{i\alpha}\epsilon_{\alpha\beta\gamma}u^{\beta}v^{\gamma}
,
\\&&
  \Braket{\big}{v}{w} =I(\alpha)\delta_{\alpha\beta}v^{\alpha}w^{\beta}
,
\end{eqnarray}
where $u$, $v$, and $w$ are the angular velocities about the principal axis and $\delta$, $\epsilon$, and $I$ are Kronecker's delta, the Levi-Civita symbol, and the principal moment of inertia, respectively.
Their combination becomes
\begin{eqnarray}
  \Braket{\big}{
    \LieBracket{}{u}{v}
  }{w}
  = I(\alpha)\epsilon_{\alpha\beta\gamma}u^{\beta}v^{\gamma}w^{\alpha}
,
\nonumber
\end{eqnarray}
which is analogous to Eq. (\ref{MC=LT}).
Therefore, this system is an example of a system that has a mathematical structure close to that of the HD, MHD, and HMHD systems.

Conversely, the KdV equation is known to be a dynamical system in the Virasoro-Bott group, whose Lie bracket and inner product are given by\cite{Michor1998}
\begin{eqnarray}
&&\nonumber
  \LieBracket{\big}{(f,a)}{(g,b)}
  =\left(fg'-f'g,\int_{S^1}f'(x) dg'(x)\right)
,
\\&&\nonumber
  \Braket{\big}{(f,a)}{(g,b)} =\int_{S^1}f(x)g(x) dx + ab
.
\end{eqnarray}
Their combination becomes
\begin{eqnarray}
&&
  \Braket{\big}{
    \LieBracket{}{(f,a)}{(g,b)}
  }{(h,c)}
  = \int_{S^1}g\left(2fh'+f'h+cf'''\right)dx
\nonumber
\end{eqnarray}
and, seemingly, cannot be rewritten as a cyclic permutation of the three modes $(f,a)$, $(g,b)$, and $(h,c)$.
This illustrates that the Lie bracket and the quadratic inner product appear necessary but are not sufficient for the structure given by Eq. (\ref{MC=LT}).

Therefore, it is possible that the remarkable relationship between the Riemannian metric, the structure constant of Lie algebra, and the eigenvalue of the relabeling operator, Eq. (\ref{MC=LT}), defines a universal sub-class of dynamical systems in Lie groups with quadratic nonlinearity.

\begin{acknowledgments}
The author expresses appreciation to 
Prof. H. Miura for his continuous encouragement
and
Prof. M. Furukawa for valuable comments.
This work was performed under the auspices of 
the NIFS Collaboration Research Program 
(NIFS13KNSS044, NIFS15KNSS065, NIFS17KNSS088, NIFS18KNSS107) 
and 
KAKENHI (Grant-in-Aid for Scientific Research(C)) 23540583, 17K05734.
The author would like to thank Enago for the English language review.
\end{acknowledgments}

\bibliography{../HMHDnotes}

\end{document}